\documentclass[article]{article}

\usepackage{hyperref}
\usepackage{graphicx}
\usepackage{color}
\usepackage{amsmath,amssymb}
\usepackage{bm}
\usepackage{mathrsfs}
\usepackage[margin=1in]{geometry}
\usepackage[sort&compress, numbers]{natbib}
\usepackage{caption}
\usepackage{subcaption}
\usepackage{verbatim}

\begin{document}

\title{Influence of constraints on axial growth reduction of cylindrical Li-ion battery electrode particles}
\author{Jeevanjyoti Chakraborty\footnote{Corresponding author. Tel: +44-1865-615144; E-mail address: \texttt{jeevanjyoti4@gmail.com}, \texttt{chakraborty@maths.ox.ac.uk}}, Colin P. Please, Alain Goriely, S. Jonathan Chapman \\ Mathematical Institute, University of Oxford, Oxford, OX2 6GG, UK}

\maketitle
\begin{abstract}
Volumetric expansion of silicon anode particles in a lithium-ion battery during charging may lead to the generation of undesirable internal stresses. For a cylindrical particle such growth may also lead to failure by buckling if the expansion is constrained in the axial direction due to other particles or supporting structures. To mitigate this problem, the possibility of reducing axial growth is investigated theoretically by studying simple modifications of the solid cylinder geometry. First, an annular cylinder is considered with lithiation either from the inside or from the outside. In both cases, the reduction of axial growth is not found to be significant. Next, explicit physical constraints are studied by addition of a non-growing elasto-plastic material: first, an outer annular constraint on a solid silicon cylinder, and second a rod-like inner constraint for an annular silicon cylinder. In both cases, it is found that axial growth can be reduced if the yield stress of the constraining material is significantly higher than that of silicon and/or the thickness of the constraint is relatively high. Phase diagrams are presented for both the outer and the inner constraint cases to identify desirable operating zones. Finally, to interpret the phase diagrams and isolate the key physical principles two different simplified models are presented and are shown to recover important qualitative trends of the numerical simulation results.
\end{abstract}


\section{Introduction}

The lithium-ion battery (LIB) has established itself as the power source of choice for mobile phones, laptop computers, and a variety of hand-held electronic devices \cite{2001NatureTarasconArmand, 2011BookFletcher, 2012ProcIEEEWhittingham}. The reason for this wide-spread use of LIBs is that they are lightweight (primarily because lithium is the lightest metal) and they have large storage capacity (because they have relatively high energy density). Motivated by this success, LIBs have been put forward, for over a decade, as the best candidates for use in electric transportation systems. Such a vision has materialized in the recent years with the launch of a number of electric-hybrid and electric vehicles. While the portable electronic devices need just one or, at most, a few unit cells in a LIB, the battery pack in a single electric car may use as many as 6000 cells \cite{2013PatentRawlinson}. The development of these cars has, therefore, resulted in a tremendous increase in the demand for LIBs. Yet, currently, wide-spread use of electric vehicles is limited by the energy capacity limits. 

The encouraging, albeit limited, success with electric cars in recent years has fostered hope that the automotive industry can indeed be driven towards a paradigm shift from a singular dependence on fossil fuels to a cleaner power source like LIBs. However, if this initial success is to be sustained and even the conservative predictions of growth trajectory for wide-spread commercial deployment are to be met, then LIBs need to have significantly higher energy capacities than current commecial ones. Only then prices may be brought down by using fewer cells in battery packs, and mileages may be improved -- thus requiring fewer charging stations. For these developments to actually happen, the current materials in commercial LIBs need to be replaced with those that, through a fundamentally different chemistry, can provide higher limits on lithium storage. For the anode, the best candidate to emerge with such a property is silicon. 

In the fully lithiated state, a single atom of Si can accommodate up to 4.4 atoms of Li resulting in the equilibrium (amorphous) phase, Li$_{22}$Si$_5$ \cite{1964KrisGlad}. This is a significant improvement on the traditional anodic material, graphite, which gives LiC$_6$ in the fully lithiated state. This translates into a theoretical specific capacity value of 4200 mAhg$^{-1}$ compared to only 372 mAhg$^{-1}$ for graphite. A higher energy capacity follows directly from this higher specific capacity.

Despite the exciting promise of such new chemistry, the use of silicon in a LIB comes with its own big challenge: lithiation of silicon results in significant volume change, which can be as high as 310\% in the fully lithiated state \cite{2001ECMSSLBeaulieu}. Slow diffusion of Li in Si results in a spatially non-uniform Li concentration, leading to stress generation due to differential growth. A similar situation arises during delithation. Indeed, it has been found that cyclic charging and discharging causes pulverization of the Si anode particles which ultimately results in a decay of the specific capacity \cite{2007JPowerSourcesKasavajjula, 2008NatureNanotechCui, 2013ScienceEbner}. This problem may be overcome to some extent by using nanostructured Si anode particles \cite{2008NatureNanotechCui, 2009NanoLettCui, 2010NanoLettSong, 2011ScienceKovalenko, 2013JMaterChemAZamfir, 2014JPhysChemLettSong, 2014AdvEnergyMaterSu}. 

\begin{figure}[ht!]
\centering
\includegraphics[width=0.5\textwidth]{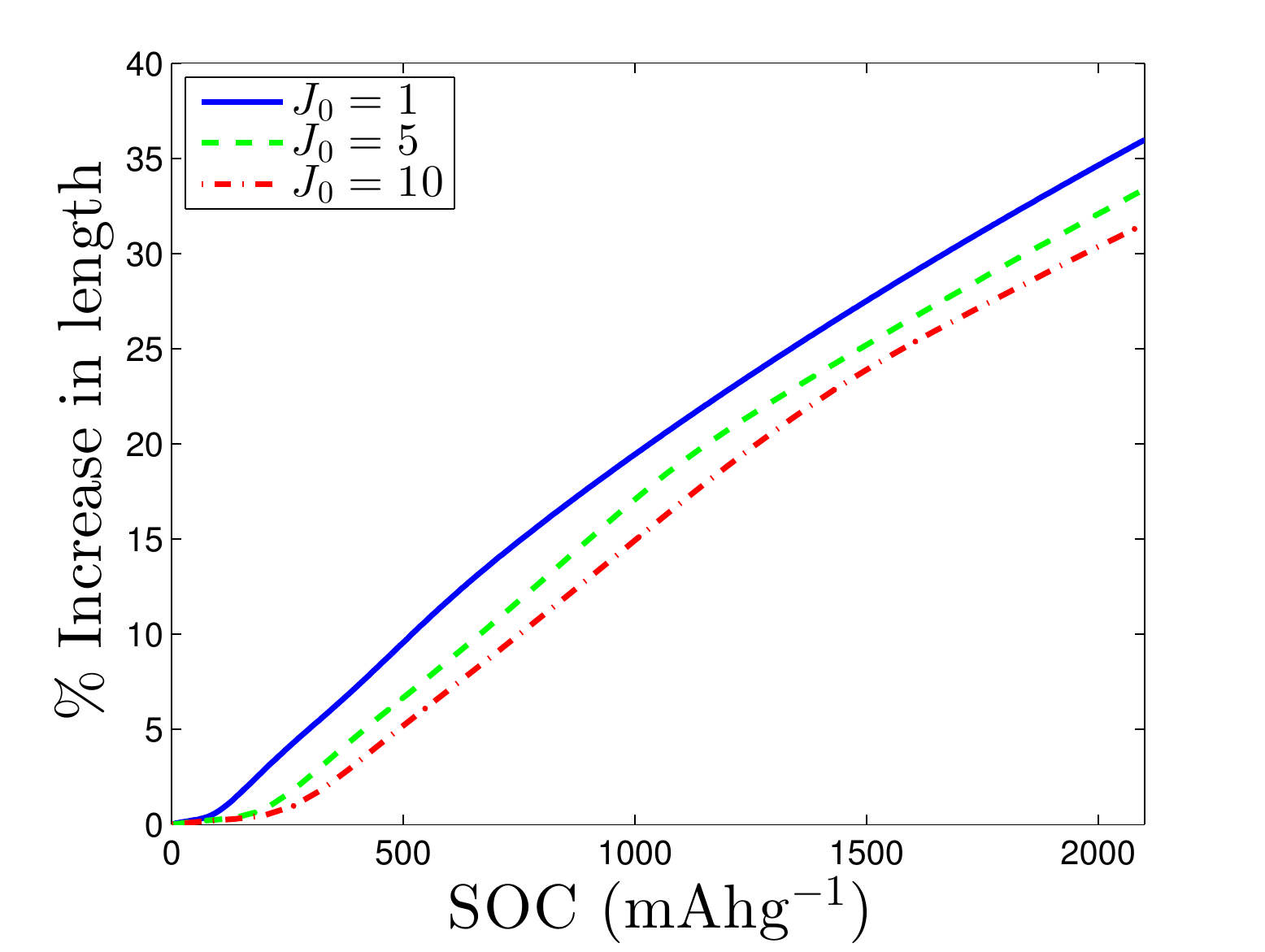}
\caption{Increase in length of an unconstrained cylindrical electrode particle with increasing state of charge for various lithiation rates (non-dimensionalized values). Values of all material properties and parameters are given in \cite{2014IJSSJeevanAdvisors}.} \label{fig:cyl}
\end{figure}

Nevertheless, even with nanostructured particles, practical challenges persist. Although the Si anode particles do not pulverize, they can still grow. This growth may be problematic within the finite confines of the electrode (and, by extension, the whole battery) packaging. While it might seem that the immediate solution to this problem would be to allow for some ``free" space within the packaging to allow for such volumetric expansion, in practice, this approach is untenable because it leads to loss of electrical contact within the electrode, and, thus to a severely impaired battery performance. Furthermore, these anode particles will be in physical contact with other particles and supporting substrates. Expansions in the presence of such constraints will lead to stresses and, possibly, mechanical failure even without fracture. Such a situation was investigated in \cite{2014IJSSJeevanAdvisors} where we studied the possibility of a cylindrical Si anode particle failing through buckling (under axial confinement). Left unconstrained, however, such a cylindrical particle grows both axially and radially during lithiation. The percentage increase in length of such a cylinder with increasing state of charge, at different charging rates, is shown in Fig.~\ref{fig:cyl}. It is this increase in length which provides the motivation for our present study.

We observe in Fig.~\ref{fig:cyl} that the percentage increases in length for three different influx rates (represented by ${J}_0$) are large and nearly equal to unconstrained, isotropic growth. Our objective here is to investigate the simplest modifications in geometry on a cylinder which might reduce axial growth. We first study the case of an annular cylinder to check if introducing a ``free space" within the particle might be useful for limiting axial growth. Next, guided by the observation that a slight spatial non-uniformity in Li concentration as seen for higher influx rates leads to a decrease in the axial growth, we impose two simple constraints in the radial direction. The question we ask ourselves is: \emph{If spatial variations in lithiation can reduce the axial growth through the influence of ``saturated" regions which do not grow any more, should it not be possible to achieve even greater reductions if we use explicit constraints?} Such constraints would be made of other materials having negligible volumetric expansion compared to that of silicon. 

The remainder of the paper is organized as follows. In Sec.~\ref{sec:formulation}, we recapitulate the formulation (in non-dimensional form) presented in our previous work \cite{2014IJSSJeevanAdvisors} for a solid Si cylinder, adapting it for the general composite cylindrical geometry considered here; the specific geometries we study in the following sections may be obtained from this as special cases. Notably, we incorporate the two-way coupling between stress and concentration of Li in Si -- meaning, that we account for the effects of both diffusion-induced stress and stress-enhanced diffusion. Our framework, while similar in spirit to a number of recent works \cite{2010JElectrochemSocSethuraman, 2011JMechPhysSolidsBower, 2011JAmCeramSocZhao, 2011JAPGaoZhou, 2011JPowerSourcesHaftbaradaran, 2012JMechPhysSolidsAnand, 2012ActaMechSinicaGao, 2013SciRepLevitas, 2013JPhysDSong, 2014JMechPhysSolidsBucci, 2014JAMGuo} involving similar two-way coupling, is, however, based on the stress-dependent chemical potential presented in \cite{2012JMechPhysSolidsCui} for spherical particles. Additionally, we also for the possibility of plastic deformations in our formulation as an important departure from prior works on cylindrical Si particles. With this framework, we are able to study high charging rates which induce stresses strong enough to reach the yield stress of Si. We relegate the specification of the boundary conditions to each specific case described in the remaining sections. In Sec.~\ref{sec:anncyl}, we discuss an annular cylinder of silicon with no explicit constraints examining two situations: lithiation occurs (i) only from the inner surface and (ii) only from the outer periphery. In Sec.~\ref{sec:outcons}, we discuss the case when a solid cylinder is constrained by an outer shell, examining the influence of the shell thickness and yield stress on axial growth. In Sec.~\ref{sec:incons}, we discuss the case of an annular cylinder constrained from the inside by a concentric cylindrical rod, again examining the influence of the radius and yield stress (of the rod) on the axial growth. In Sec.~\ref{sec:phase_diag}, we present phase-diagrams of the percentage increases in length for both the outer and the inner constraint cases corresponding to different combinations of the radius of the constraint and the ratio of the yield-stress of the constraining material to that of silicon. In Sec.~\ref{sec:sm1}, we describe our first simplified model based on a simple force balance, and highlight two scaling relationships that characterize the phase-diagrams. In Sec.~\ref{sec:sm2}, we describe our second simplified model which can recover some of the qualitative trends of the phase-diagrams. Finally, in Sec.~\ref{sec:conclusions}, we summarize our findings. 

\section{Mathematical Formulation} \label{sec:formulation}

%
\begin{figure}[h!]
\centering
\includegraphics[width=0.4\textwidth]{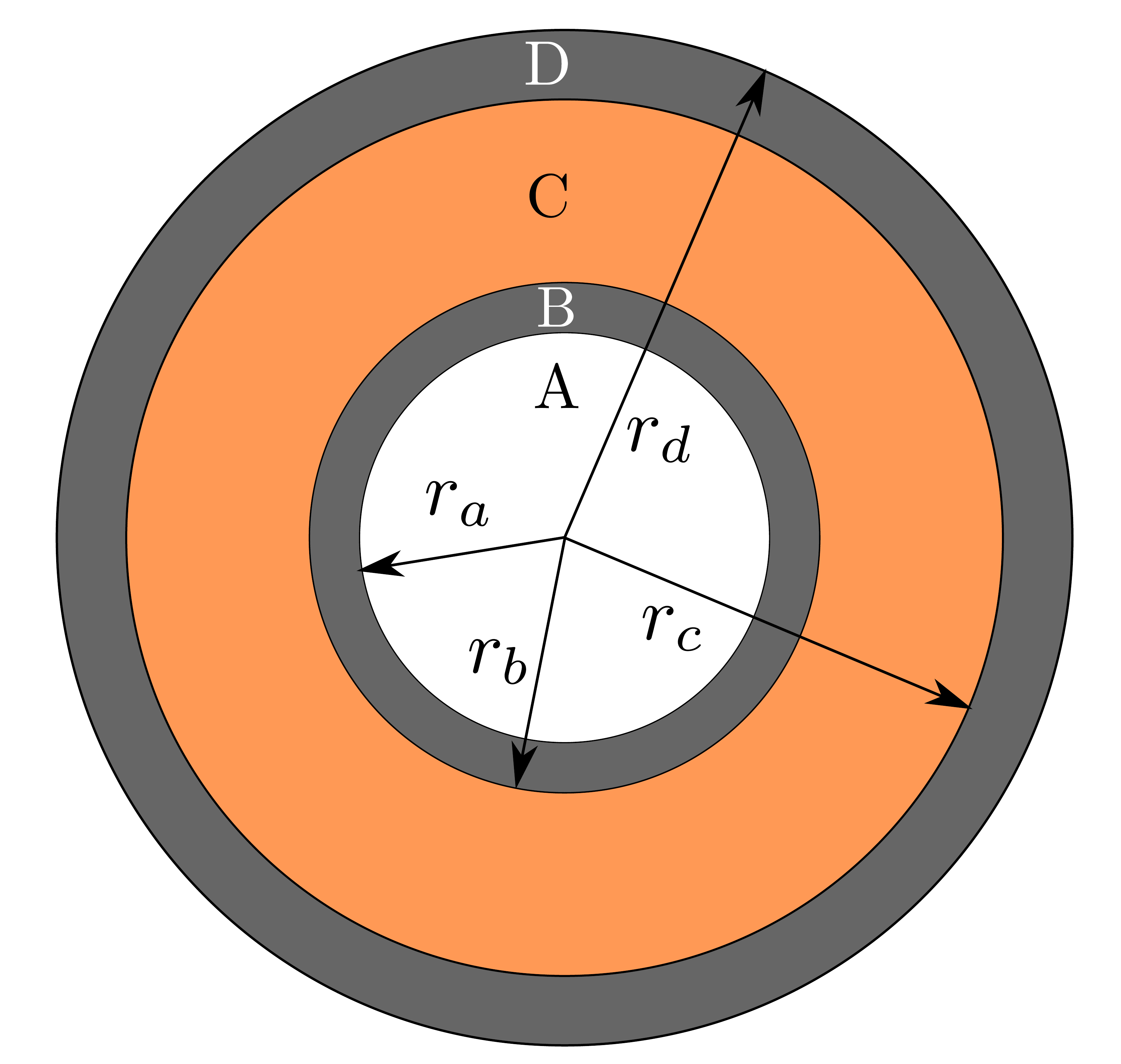}
\caption{General schematic of the silicon electrode particle with constraints.} \label{fig:general}
\end{figure}
We consider a general circular cylinder made up of a number of concentric shells as shown in Fig.~\ref{fig:general}. In terms of a non-dimensional radial coordinate variable, ${r}$, we have 
\begin{align*}
0\le {r} \le r_a & \qquad {\text{Region A: Empty region}} \\
r_a\le {r} \le r_b  & \qquad {\text{Region B: Inner constraining material}} \\
r_b\le {r} \le r_c & \qquad {\text{Region C: Silicon}} \\
r_c\le {r} \le r_d & \qquad {\text{Region D: Outer constraining material}}
\end{align*}
The only geometric restriction we impose is that the volume of silicon in Region C is equal to that in a solid circular cylinder with radius ${r}=1$, so that there is a consistent basis for comparison of the stresses and deformation induced by the lithation-driven expansion of silicon. Per unit length, this requirement translates as: 
\begin{gather}
r_c^2-r_b^2=1.
\end{gather}
We extend the formulation from our previous work on such a solid circular cylinder \cite{2014IJSSJeevanAdvisors} to this situation. Any general point in this composite cylinder is, in cylindrical coordinates, ${r} \in (0,r_d)$, $\theta \in (0,2\pi)$, $z \in (0,l)$, and the non-dimensionalized set of equations governing the lithiation-driven mechanical and chemical processes are:
\begin{align}
\frac{\partial c}{\partial {t}} & = -\frac{\partial {J}_r}{\partial {r}} - \frac{{J}_r}{{r}}, \label{eq:nondim_c}\\ 
0 &= \frac{\partial {\sigma}_r^0}{\partial {r}} + \frac{{\sigma}_r^0 - {\sigma}_\theta^0}{{r}}, \label{eq:nondim_sigma} \\
\frac{\partial \lambda_r}{\partial {t}} &= \sqrt{\frac{3}{2}}\lambda_r \dot{d}_0 \frac{R_0^2}{D_0} \left( \frac{{\sigma}_{\rm{eff}}}{{\sigma}_f} - 1\right)^m \frac{{\tau}_r}{ \sqrt{{\tau}_r^2 + {\tau}_\theta^2 + {\tau}_z^2} } \; H\left( \frac{{\sigma}_{\rm{eff}}}{{\sigma}_f} - 1 \right), \label{eq:nondim_lambdar}\\
\frac{\partial \lambda_\theta}{\partial {t}} &= \sqrt{\frac{3}{2}}\lambda_\theta \dot{d}_0 \frac{R_0^2}{D_0} \left( \frac{{\sigma}_{\rm{eff}}}{{\sigma}_f} - 1\right)^m \frac{{\tau}_\theta}{ \sqrt{{\tau}_r^2 + {\tau}_\theta^2 + {\tau}_z^2} } \; H\left( \frac{{\sigma}_{\rm{eff}}}{{\sigma}_f} - 1 \right) \label{eq:nondim_lambdath}, \\
\lambda_z &= \frac{1}{\lambda_r \lambda_{\theta}}. \label{eq:nondim_lambdaz}
\end{align}
Eq.~(\ref{eq:nondim_c}) governs the transport of Li in the material, with $c$ representing the non-dimensional measure of Li concentration, and ${J}_r$ the flux.  Eq.~(\ref{eq:nondim_sigma}) represents the mechanical equilibrium in the reference (undeformed) configuration in terms of the first Piola-Kirchhoff stresses, ${\sigma}_r^0$ and ${\sigma}_{\theta}^0$, along the radial and circumferential (or, hoop) directions, respectively. Eqs.~(\ref{eq:nondim_lambdar}) and (\ref{eq:nondim_lambdath}) govern the evolution of the plastic stretches, along the radial and hoop directions, respectively. The parameter $\dot{d}_0$ is the characteristic strain rate, $D_0$ is the concentration-independent diffusivity of silicon, $R_0$ is a reference radius, and $\sigma_f$ is the yield stress of silicon. Also, $\tau_r$, $\tau_\theta$, and $\tau_z$ are the deviatoric parts of the Cauchy stresses while $\sigma_{\rm{eff}}$ is the effective stress; these stresses are defined later. Eq.~(\ref{eq:nondim_lambdaz}) indicates that the plastic deformation is assumed to be volume-preserving (see \cite{2014IJSSJeevanAdvisors} for details). The first Piola-Kirchhoff stresses may be expressed in terms of the strains, $E_r^e$, $E_{\theta}^e$, and $E_z^e$, as
\begin{subequations} \label{eq:nondim_PK-strains}
\begin{align}
{\sigma}_r^0 &= J^c \frac{{E}(c)}{(1+\nu)(1-2\nu)} \left[ (1-\nu)E_r^e + \nu \left( E_\theta^e + E_z^e  \right)  \right] \frac{2E_r^e + 1}{1+\partial {u}/\partial {r}}, \\
{\sigma}_\theta^0 &= J^c \frac{{E}(c)}{(1+\nu)(1-2\nu)} \left[ (1-\nu)E_\theta^e + \nu \left( E_z^e + E_r^e  \right)  \right] \frac{2E_\theta^e + 1}{1+ {u}/{r}}, \\
{\sigma}_z^0 &= J^c \frac{{E}(c)}{(1+\nu)(1-2\nu)} \left[ (1-\nu)E_z^e + \nu \left( E_r^e + E_\theta^e  \right)  \right] \frac{2E_z^e + 1}{1+\partial {w}/\partial {z}}.
\end{align}
\end{subequations}
Here, $u$ and $w$ are the radial and the axial displacements respectively. Further, $J^c=1+3\eta x_{\rm{max}}c$ represents the volumetric expansion of Si or the constraining material due to lithiation, ${E}(c)=1+\eta_Ex_{\rm{max}}c$ is the concentration-dependent non-dimensional modulus of elasticity, and $\nu$ is Poisson's ratio. Note that the saturation level of Li in Si is set by the factor, $x_{\rm{max}}=4.4$. In the constraining material (of which Regions B and D are made), this saturation level is set at an arbitrarily small value of $x_{\rm{max}} = 4.4 \times 10^{-3}$. It is this difference in the saturation levels which results in a lower volumetric expansion of Regions B and D compared to Region C, and imparts their functionality as external physical constraints against the expansion of silicon (since radial displacements at the interfaces of Regions B, C, and D must match). The parameter $\eta$ represents the coefficient of compositional expansion while $\eta_E$ represents the rate of change of modulus of elasticity with concentration; these values are taken to be the same for both silicon and the constraining material. 

The stresses may be linked to the displacement by expressing the strains in terms of the displacement as
\begin{subequations} \label{eq:nondim_strain-displacement}
\begin{align}
E_r^e &= \frac{1}{2} \left[ \left( F_r^e \right)^2 - 1 \right] = \frac{1}{2} \left(J^c\right)^{-2/3} \frac{\left( 1+\partial {u}/\partial {r} \right)^2}{\lambda_r^2} - \frac{1}{2}, \\
E_\theta^e &= \frac{1}{2} \left[ \left( F_\theta^e \right)^2 - 1 \right] = \frac{1}{2} \left(J^c\right)^{-2/3} \frac{\left( 1+{u}/{r} \right)^2}{\lambda_\theta^2} - \frac{1}{2}, \\
E_z^e &= \frac{1}{2} \left[ \left( F_z^e \right)^2 - 1 \right] = \frac{1}{2} \left(J^c\right)^{-2/3} \frac{\left( 1+\partial {w}/\partial {z} \right)^2}{\lambda_z^2} - \frac{1}{2}.
\end{align}
\end{subequations}
In Eq.~(\ref{eq:nondim_c}) the non-dimensional flux is given by
\begin{align} 
{J}_r = - {D} c \frac{\partial {\mu}}{\partial {r}}, \label{eq:nondim_flux}
\end{align}
where
\begin{gather}
{\mu} = \frac{\mu_0^0}{R_g T} + \log(\gamma c) + {\mu}_{S1} + {\mu}_{S2} + {\mu}_{S3},
\end{gather}
with
\begin{subequations}
\begin{align}
{\mu}_{S1} &=  -\frac{1}{6 x_{\rm{max}}} \frac{\partial J^c}{\partial c} \frac{{E}}{(1+\nu)(1-2\nu)} \left[ (1-\nu) \left\{ (E_r^e)^2 + (E_\theta^e)^2 + (E_z^e)^2 \right\} \right. \nonumber \\
& \left. \qquad +2\nu (E_r^e E_\theta^e + E_\theta^e E_z^e + E_z^e E_r^e)  \right], \\
{\mu}_{S2} &= -\frac{1}{3 x_{\rm{max}}} \frac{\partial J^c}{\partial c} \frac{{E}}{(1+\nu)(1-2\nu)} \left[ (1+\nu) (E_r^e + E_\theta^e + E_z^e) \right], \\ 
{\mu}_{S3} &= \frac{1}{2 x_{\rm{max}}} J^c \left[ \frac{\partial}{\partial c} \left\{ \frac{{E}(1-\nu)}{(1+\nu)(1-2\nu)} \right\} \left\{ (E_r^e)^2 + (E_\theta^e)^2 + (E_z^e)^2  \right\} \right. \nonumber \\ 
& \qquad \left. + 2 \frac{\partial}{\partial c} \left\{ \frac{{E}\nu}{(1+\nu)(1-2\nu)}  \right\} (E_r^eE_\theta^e + E_\theta^eE_z^e + E_z^eE_r^e)  \right].
\end{align}
\end{subequations}
Eqs.~(\ref{eq:nondim_lambdar}) and (\ref{eq:nondim_lambdath}) are expressed in terms of the non-dimensionalized deviatoric parts of the Cauchy stress tensor given by
\begin{align}
{\tau}_{r,\theta,z} = {\sigma}_{r,\theta,z} - \frac{1}{3}({\sigma}_r + {\sigma}_\theta + {\sigma}_z),
\end{align}
where
\begin{subequations} \label{eq:nondim_Cauchy}
\begin{align}
{\sigma}_r &= \frac{{E}}{(1+\nu)(1-2\nu)}\left[ (1-\nu)E_r^e + \nu\left( E_\theta^e + E_z^e  \right)  \right] \frac{\sqrt{2E_r^e + 1}}{\sqrt{2E_\theta^e + 1}\sqrt{2E_z^e + 1}} \\
{\sigma}_\theta &= \frac{{E}}{(1+\nu)(1-2\nu)}\left[ (1-\nu)E_\theta^e + \nu\left( E_z^e + E_r^e  \right)  \right] \frac{\sqrt{2E_\theta^e + 1}}{\sqrt{2E_z^e + 1}\sqrt{2E_r^e + 1}} \\
{\sigma}_z &= \frac{{E}}{(1+\nu)(1-2\nu)}\left[ (1-\nu)E_z^e + \nu\left( E_r^e + E_\theta^e  \right)  \right] \frac{\sqrt{2E_z^e + 1}}{\sqrt{2E_r^e + 1}\sqrt{2E_\theta^e + 1}}.
\end{align}
\end{subequations}
Furthermore, the non-dimensional effective stress in Eqs.~(\ref{eq:nondim_lambdar}) and (\ref{eq:nondim_lambdath}) is given by
\begin{align}
{\sigma}_{\rm{eff}} = \sqrt{\frac{3}{2}} \sqrt{{\tau}_r^2 + {\tau}_\theta^2 + {\tau}_z^2}.  \label{eq:nondim_effectivestress}
\end{align}
The boundary and the initial conditions associated with these equations depend on the particular configuration under consideration, and will be described on a case by case basis in the following sections. Common to all situations, however, is the physical condition that the cylinder ends are free. This implies that there can be no axial force acting on the ends, so that
\begin{gather}
2 \pi \int_{r_{\rm{in}}}^{r_{\rm{out}}} {\sigma}_z^0  {r} \; d{r}=0,
\end{gather}
where $r_{\rm{in}}$ and $r_{\rm{out}}$ are the appropriate inner and outer limits of the radius.

Before proceeding, we note that in almost all the cases we investigate, the percentage increase in length is lower for higher values of ${J}_0$ when lithiation is from the outside (the exceptions are Figs.~\ref{fig:outcons}~(d) and \ref{fig:incons}~(d), which will be studied separately). The same trend is observed in the unconstrained solid cylinder case; see Fig.~\ref{fig:cyl}. This trend can be explained on the basis of the spatial heterogeneity of lithiation. For a relatively high influx rate, the amount of lithiation is higher in the periphery than in the inner region leading to a greater spatial heterogeneity of lithiated silicon over the cylinder radius compared to situations with a relatively low influx rate. The unlithiated region then constrains the lithiated periphery resulting in more plastic flow and lower axial expansion. The same explanation holds for annular silicon regions (irrespective of the nature of constraints) undergoing lithiation at different influx rates.

%
\begin{table}[t!] 
\caption{} \label{table:values}
\centering
\begin{tabular}{l l}
\hline \hline
Material property or parameter & Value \\
\hline
$A_0$, parameter used in activity constant & -29549 Jmol$^{-1\;a}$ \\
$B_0$, parameter used in activity constant & -38618 Jmol$^{-1\;a}$ \\
$D_0$, diffusivity of Si & 1 $\times$ 10$^{-16}$ m$^2$s$^{-1\;b}$ \\
$\dot{d}_0$, characteristic strain rate for plastic flow in Si & 1 $\times 10^{-3}$ s$^{-1\;a}$ \\
$E_0$, modulus of elasticity of pure Si & 90.13 GPa$^{\;a}$ \\
$m$, stress exponent for plastic flow in Si & 4$^{\;c}$ \\
$R_g$, universal gas constant & 8.314 JK$^{-1}$mol$^{-1}$ \\
$R_0$, initial radius of unlithiated Si electrode & 200 nm \\
$T$, temperature & 300 K \\
$V_m^B$, molar volume of Si & 1.2052 $\times$ 10$^{-5}$ m$^3$mol$^{-1\;a}$ \\
$x_{\rm{max}}$, maximum concentration of Li in Si & 4.4 \\
$\alpha$, coefficient of diffusivity & 0.18$^{\;d}$ \\
$\eta$, coefficient of compositional expansion & 0.2356$^{\;a}$ \\
$\eta_E$, rate of change of modulus of elasticity with concentration & -0.1464$^{\;a}$ \\
$\nu$, Poisson's ratio of Si & 0.28$^{\;a}$ \\
$\sigma_f$, initial yield stress of Si & 0.12 GPa$^{\;c}$ \\
\hline \\
$^a$ \cite{2012JMechPhysSolidsCui} \\
$^b$ \cite{2011NanoLettLiu} \\
$^c$ \cite{2011JMechPhysSolidsBower} \\
$^d$ \cite{2011JPowerSourcesHaftbaradaran}
\end{tabular}
\end{table}

In what follows we look at special cases, focussing on the percentage increase in length that results from lithiation up to a 50\% state of charge (SOC) for various values of the lithation rates. The motivation for choosing this particular value of the SOC is three-fold. First, in an actual battery, the anode particles which are farthest from the separator (nearest the current collector) might not reach a 100\% lithiated state. So, a 50\% SOC is representative for the entire collection of particles. Second, as the SOC builds up, there is increased resistance to further charging, and so going up to a 100\% SOC is more time-consuming for our numerical simulations. Third, if the axial growth corresponding to 50\% SOC is practically untenable, it certainly is so for higher values of the SOC. Thus, although our findings are based on the 50\% SOC, our prescriptions to overcome undesirable scenarios are sufficiently general to be usefully applied for any value of the SOC. Unless otherwise stated, we use the values of the material properties and the parameters (taken from \cite{2014IJSSJeevanAdvisors}) in Table~\ref{table:values}. The constraint (either the inner or the outer) and the silicon regions are distinguished primarily by a difference in the volumetric expansion, and, in particular situations, also by a difference in the yield stress value. 

\section{Annular cylinder} \label{sec:anncyl}

To realize the annular cylinder geometry, we let the Regions B and D vanish, and retain Region C. Thus, $r_{\rm{in}}=r_b \le {r} \le r_c=r_{\rm{out}}$. We consider two sub-cases: (a) lithiation only from the inside, and (b) lithiation only from the outside. The initial and boundary conditions corresponding to Eq.~(\ref{eq:nondim_c}) are
\begin{align}
c({r},0) &=0,
\end{align}
\begin{align}
&\text{(a) Lithiation only from inside:  } {J}_r(r_b,{t}) = {J}_0(1-c), \quad {J}_r(r_c,{t}) =0, \\
&\text{(b) Lithiation only from outside:  } {J}_r(r_b,{t}) = 0, \quad {J}_r(r_c, {t}) = {J}_0(1-c). 
\end{align}
For both (a) and (b), the boundary conditions for the stresses are 
\begin{gather}
{\sigma}_r^0(r_b,{t}) = 0, \quad {\sigma}_r^0(r_c,{t}) = 0.
\end{gather}
Again for both (a) and (b), the initial conditions corresponding to Eqs.~(\ref{eq:nondim_lambdar}) and (\ref{eq:nondim_lambdath}) are:
\begin{gather}
\lambda_r({r},0)=1, \quad \text{and} \quad \lambda_\theta({r},0)=1.
\end{gather}
%
\begin{figure}[ht!]
\centering
\begin{subfigure}[b]{0.5\textwidth}
\includegraphics[width=\textwidth]{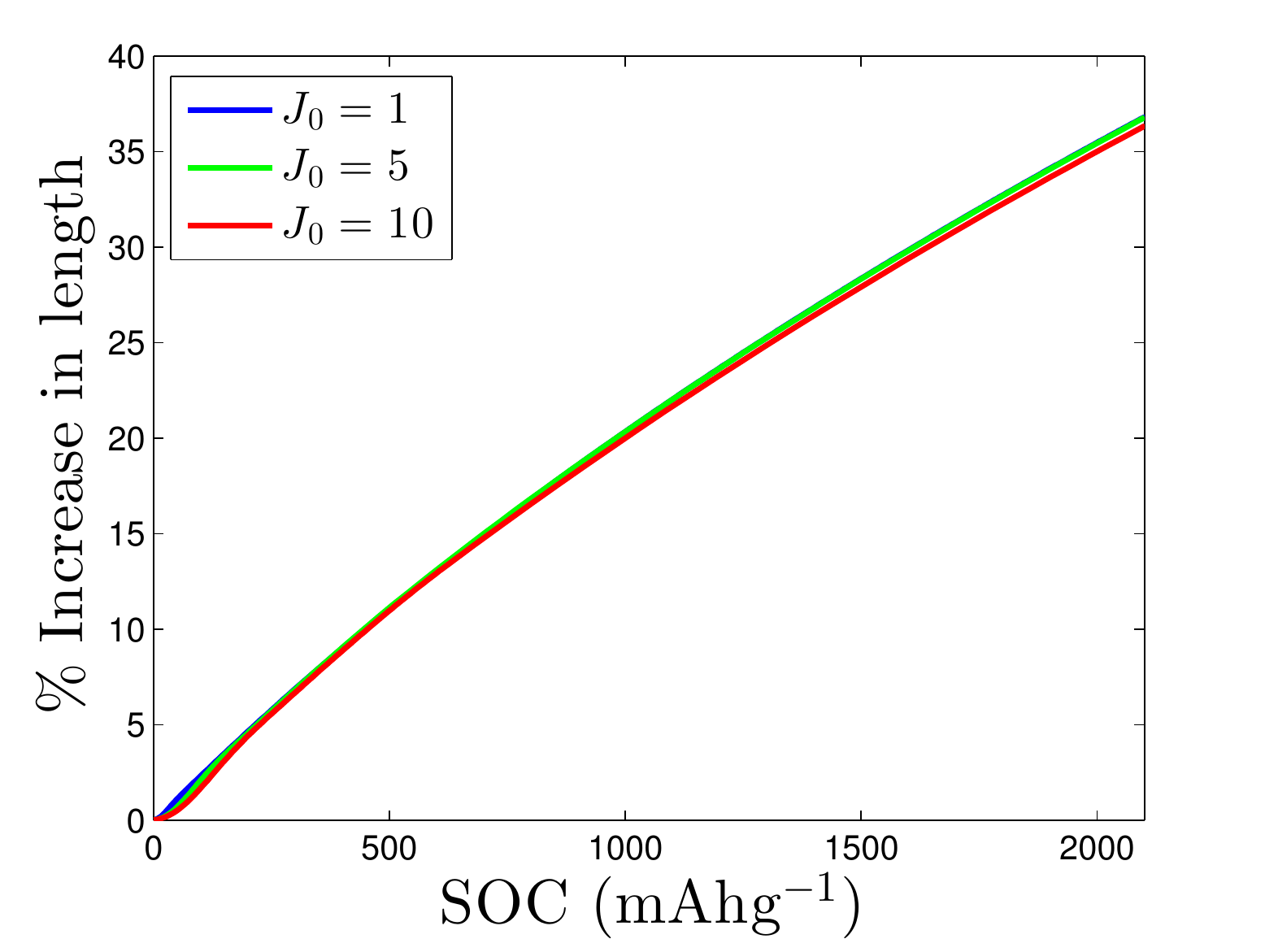}
\caption{}
\end{subfigure}%
\begin{subfigure}[b]{0.5\textwidth}
\includegraphics[width=\textwidth]{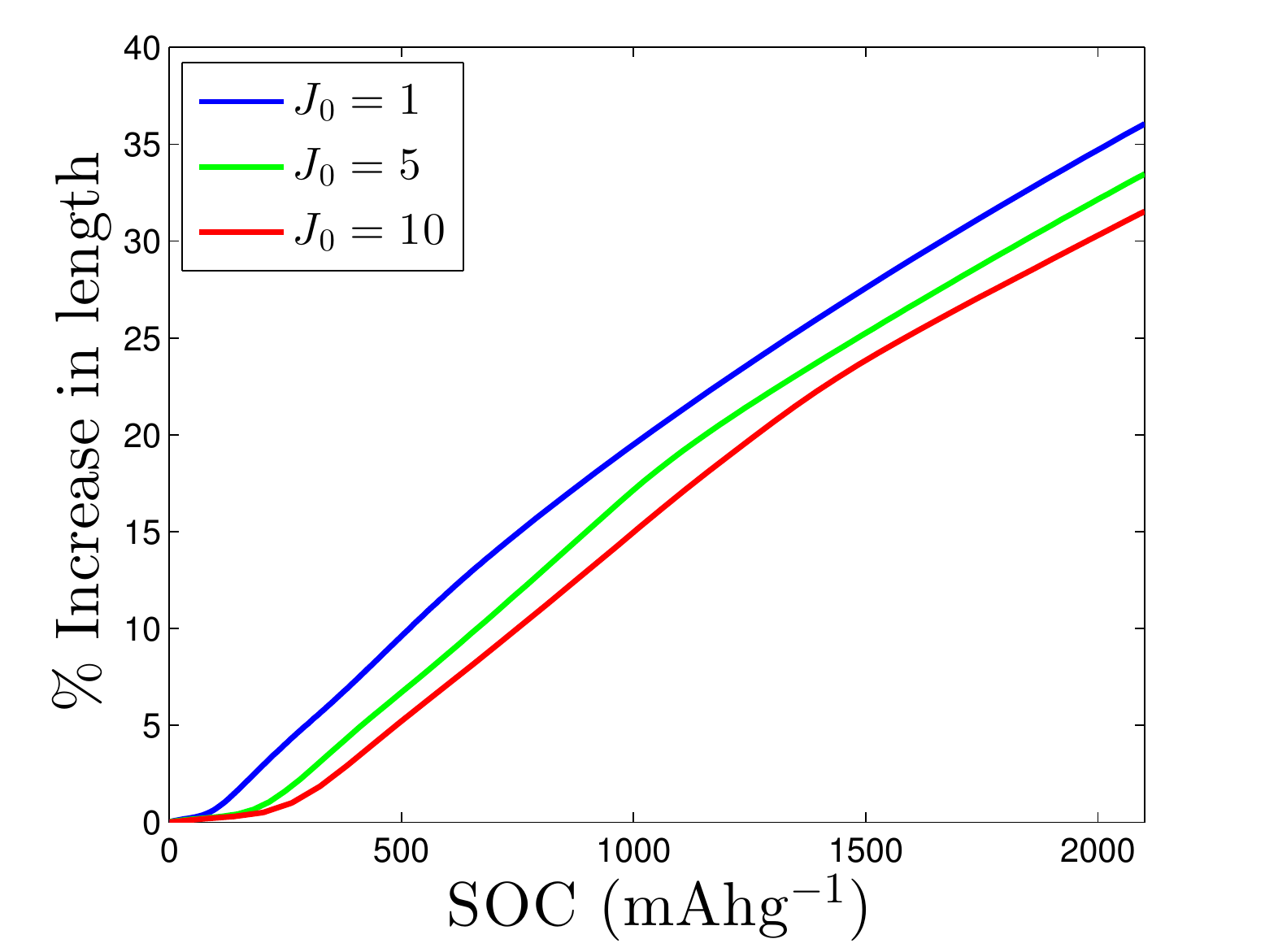}
\caption{}
\end{subfigure}
\caption{Increase in length of an annular electrode particle with increasing state of charge for various lithiation rates (non-dimensionalized values). Lithiation from (a) inside and (b) outside.} \label{fig:annular}
\end{figure}

The predicted percentage increase in length corresponding to the two cases are shown in Fig.~\ref{fig:annular}~(a) and (b), where $r_b=0.1$. We choose this particular value of $r_b$ so that the radius of the annulus hole is significantly smaller than the silicon thickness. We observe that the percentage increase in length for different influx rates in (a) are independent of the charge rate unlike in (b) where such values are lower for higher influx rates. The reason for this trend is that strong hoop stresses are generated at the inner boundary of the annulus during lithiation which results in a higher value of the diffusion coefficient. However, since in (a) the influx is from the inner boundary near which diffusion is strong, there is greater spatial homogeneity of lithiated silicon. In (b) even though diffusion is faster near the inner boundary, spatial heterogeneity results at the outer boundary resulting in a decrease in axial extension. 

We make two important remarks. First, the percentage increase in length when $r_b=0.5$ is almost the same as that when $r_b=0.1$. Since increasing the value of $r_b$ also implies increasing the value of $r_c$ to preserve the silicon content, there is no practical benefit of having an annulus with larger inner radius. Second, we have also examined the case when the radial displacement at the outer periphery is constrained to be zero so that all radial growth is accommodated by filling in the hole of the annulus. However, such a set-up leads to extremely high ($> 100\%$) increases in length, and is, thus, antithetical to the overall objective of this investigation.

\section{Outer constraint} \label{sec:outcons}

We let Regions A and B vanish so that $r_b=0$, and we have again a solid circular cylinder of silicon; Region C is thus defined by $r_{\rm{in}}=r_b = 0 \le {r}\le1=r_c$. The outer constraint (Region D) is retained; $r_{\rm{out}}=r_d$. Lithiation is considered to occur from the outside through Region D.  The initial and boundary conditions are
\begin{gather}
c({r},0) =0, \\
{J}_r(0,{t}) = 0, \quad {J}_r(r_d, {t}) = {J}_0(1-c), \\
{u}(0,{t}) = 0, \quad {\sigma}_r^0(r_d,{t}) = 0,
\lambda_r({r},0)=1, \quad \text{and} \quad \lambda_\theta({r},0)=1.
\end{gather}
%


\begin{figure}[ht!]
\centering
\begin{subfigure}[b]{0.5\textwidth}
\includegraphics[width=\textwidth]{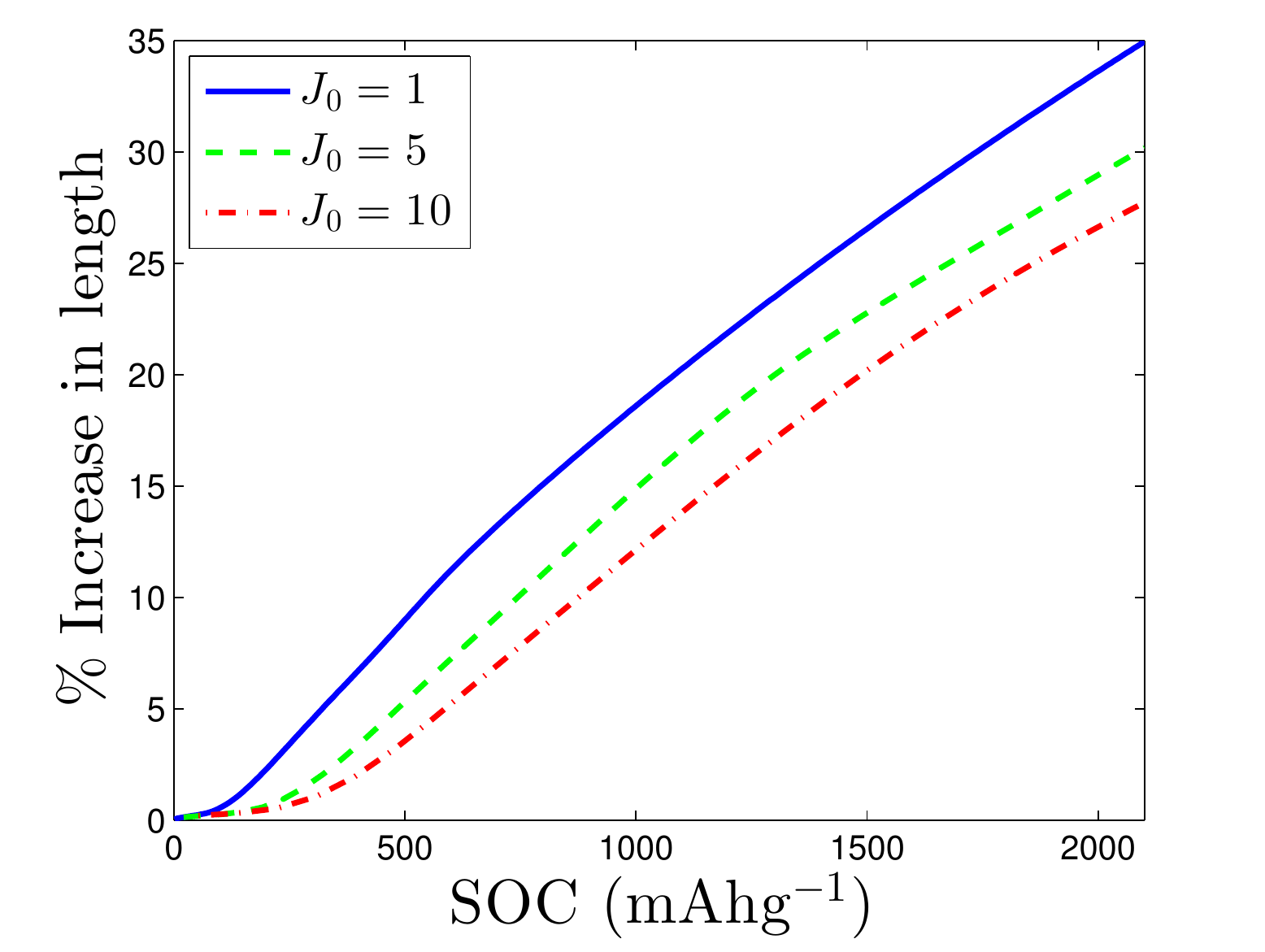}
\caption{}
\end{subfigure}%
\begin{subfigure}[b]{0.5\textwidth}
\includegraphics[width=\textwidth]{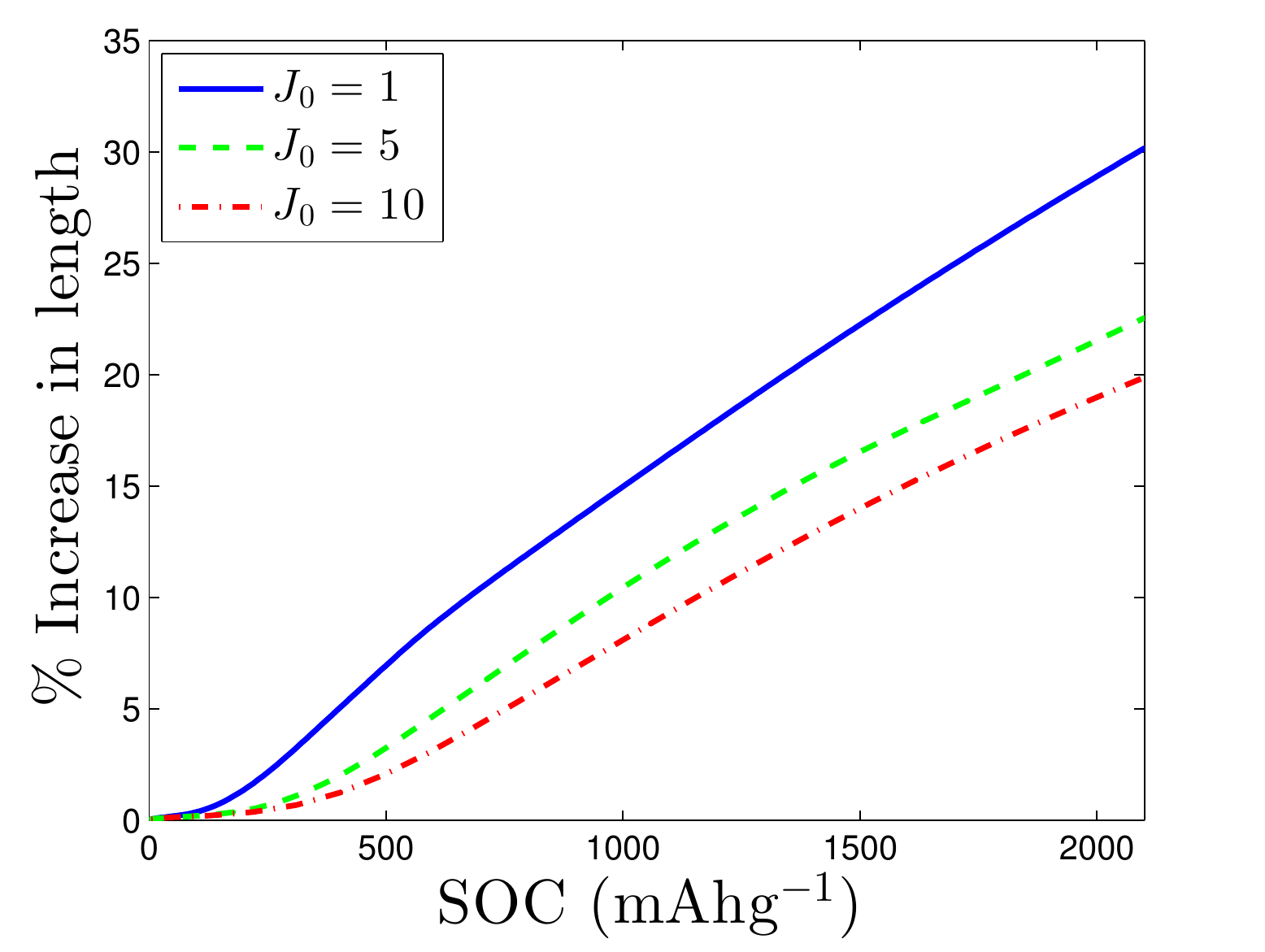}
\caption{}
\end{subfigure}
\begin{subfigure}[b]{0.5\textwidth}
\includegraphics[width=\textwidth]{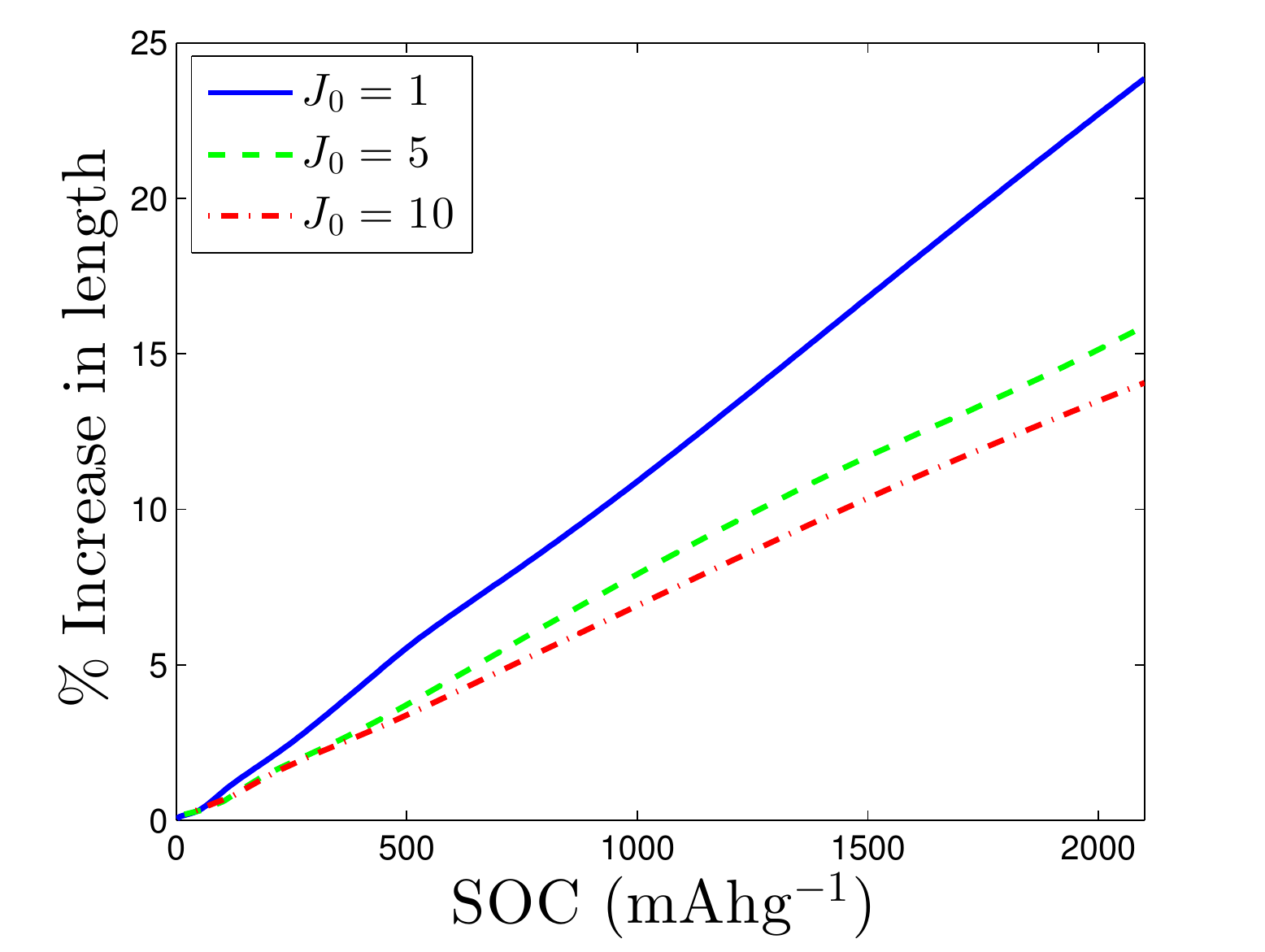}
\caption{}
\end{subfigure}%
\begin{subfigure}[b]{0.5\textwidth}
\includegraphics[width=\textwidth]{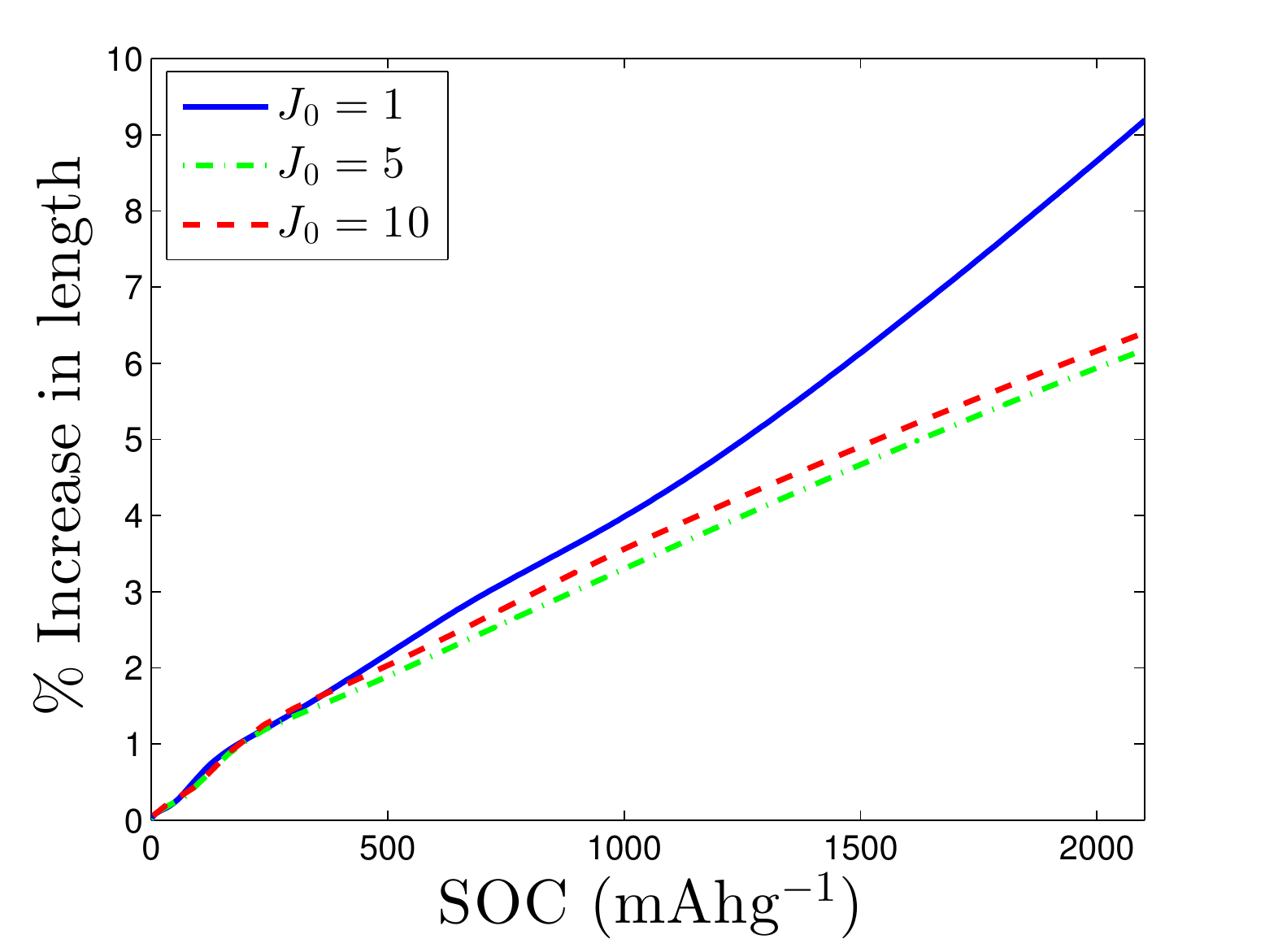}
\caption{}
\end{subfigure}
\caption{Increase in length of a cylindrical electrode particle constrained from the outside for various lithation rates (non-dimensionalized values). Interface between constraining outer rod and cylindrical particle is at ${r} = r_c = 1$. Outer radius of the constraining annulus is at ${r}= r_d = 1.1$ in (a) and (c), and at ${r}=r_d=1.3$ in (b) and (d). Yield stress of constraining annulus (Region D) is the same as that of the silicon cylinder (Region C) in (a) and (b), and 10 times higher than that of cylinder in (c) and (d).} \label{fig:outcons}
\end{figure}

We now study the influence of the thickness and the yield stress of the constraining material on the percentage increase in length. We choose two different thicknesses such that the outer radius becomes $r_d=1.1$ and $r_d=1.3$. The choice of $r_d=1.1$ is motivated by the desire to study the case where the constraining material thickness is one order of magnitude smaller than the radius of the silicon region; on the other hand, $r_d=1.3$ is chosen arbitrarily to look at the influence of a considerably larger thickness. Fig.~\ref{fig:outcons} shows the predicted percentage increase in length for $r_d=1.1$ in (a) and (c), and for $r_d=1.3$ in (b) and (d). In (a) and (b) the yield stress value in Region D is the same as in Region C while in (c) and (d) it is 10 times higher in Region D than that in Region C. 

We observe that an increase in $r_d$ reduces the percentage increase in length. The reason for this trend appears to be that Region D undergoes a lower volumetric expansion compared to Region C. However, displacements must be continuous across the interface between Regions D and C. This condition, therefore, constrains the axial deformation of Region C. Such a constraining influence, of course, gets stronger as Region D gets thicker. Additionally, increasing the yield stress value of Region D constrains the overall axial deformation even further. Thus, as seen in (d), the percentage increase in length is significantly reduced when the yield stress of Region D is 10 times higher than that of Region C \emph{and} the thickness of Region D is high. An interesting observation is that, unlike other cases, for (d), the percentage increase in length corresponding to ${J}_0=5$ is slightly smaller than that corresponding to ${J}_0=10$ for sufficiently high values of the SOC.

\section{Inner constraint} \label{sec:incons}

\begin{figure}[ht!]
\centering
\begin{subfigure}[b]{0.5\textwidth}
\includegraphics[width=\textwidth]{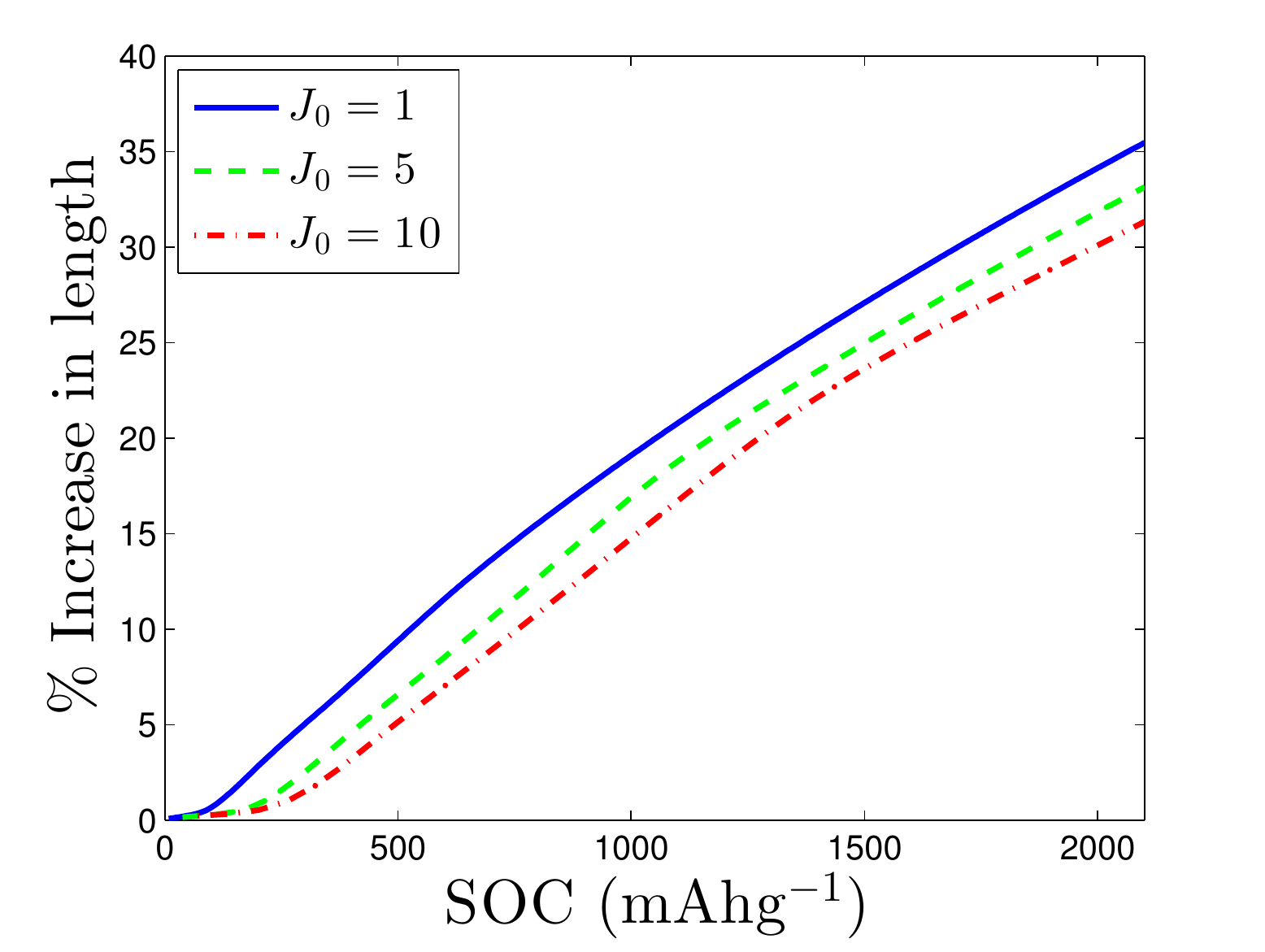}
\caption{}
\end{subfigure}%
\begin{subfigure}[b]{0.5\textwidth}
\includegraphics[width=\textwidth]{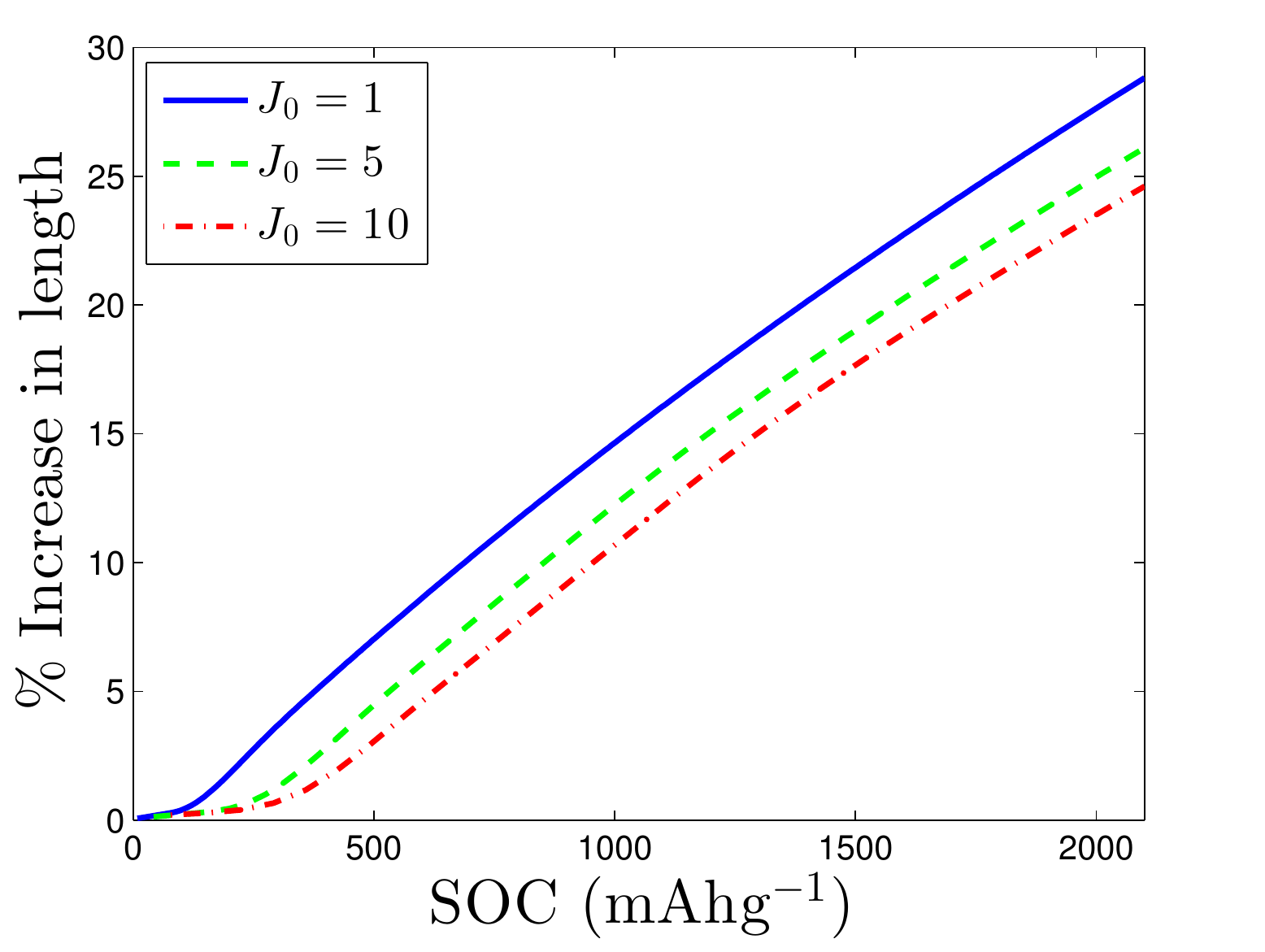}
\caption{}
\end{subfigure}
\begin{subfigure}[b]{0.5\textwidth}
\includegraphics[width=\textwidth]{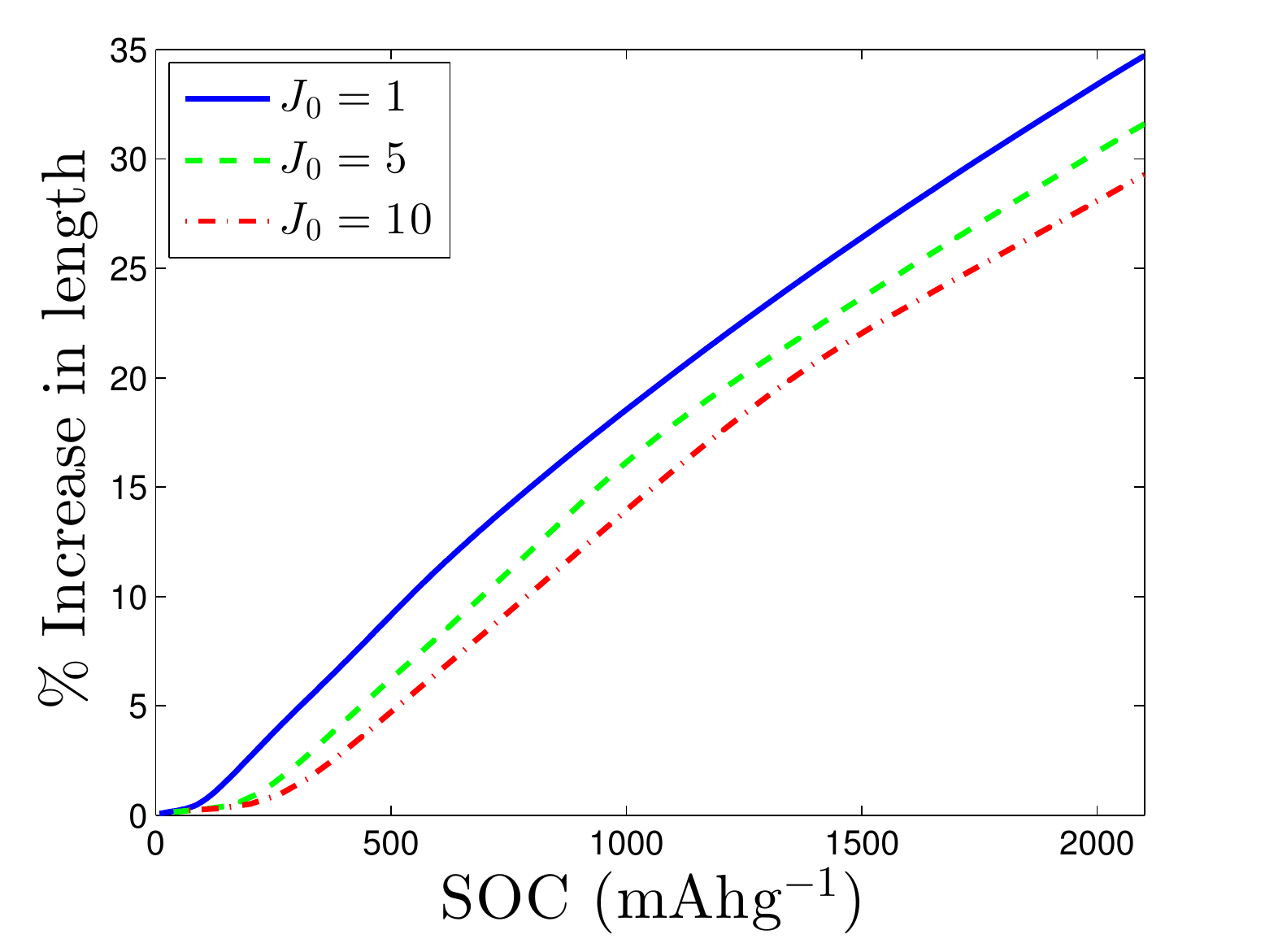}
\caption{}
\end{subfigure}%
\begin{subfigure}[b]{0.5\textwidth}
\includegraphics[width=\textwidth]{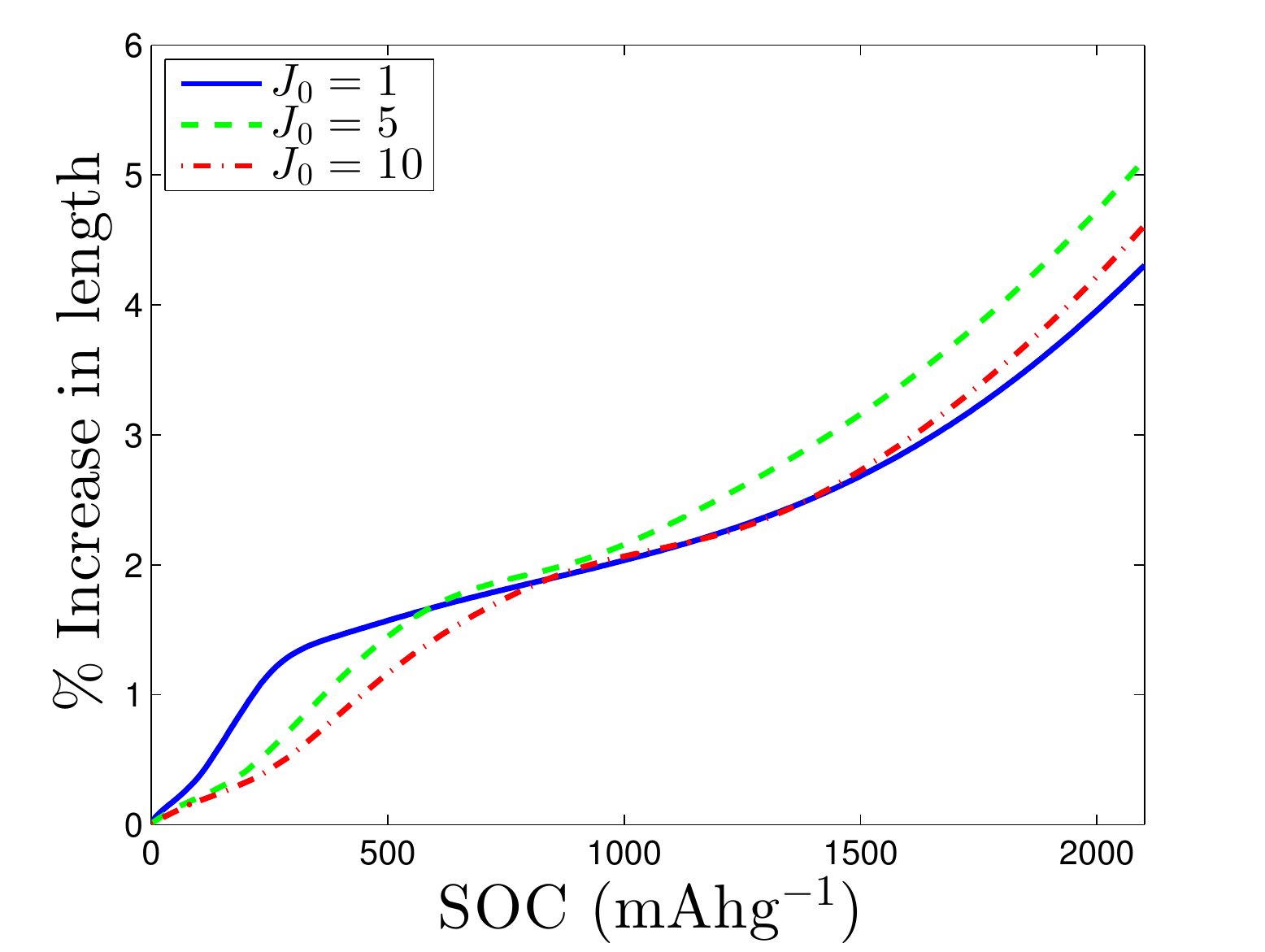}
\caption{}
\end{subfigure}%
\caption{Increase in length of an annular electrode particle constrained from the inside for various lithiation rates (non-dimensionalized values). Interface between constraining inner rod and annular particle is at ${r} = r_b = 0.1$ in (a) and (c), and at ${r} = r_b = 0.5$ in (b) and (d). Outer radius of the Si annulus is at ${r}= r_c = \sqrt{1+0.1^2}=1.005$ in (a) and (c), and at ${r}=r_c=\sqrt{1+0.5^2}=1.118$ in (b) and (d). Yield stress of inner rod (Region B) is the same as that of the annulus (Region C) in (a) and (b), and 10 times higher than that of annulus in (c) and (d).} \label{fig:incons}
\end{figure}

For this case, we let Regions A and D vanish. Thus the inner constraint is now a solid circular cylinder defined in the region, $r_{\rm{in}} = r_a = 0 \le {r} \le r_b$. Region C retains its annular shape, and is defined in the region $r_b \le {r} \le r_c = r_{\rm{out}}$. Lithiation is considered to occur only from the outside and not through Region B. The initial and boundary conditions are
\begin{gather}
c({r},0) =0, \\
{J}_r(0,{t}) = 0, \quad {J}_r(r_c, {t}) = {J}_0(1-c),
\end{gather}
%
%
\begin{gather}
{u}(0,{t}) = 0, \quad {\sigma}_r^0(r_c,{t}) = 0,
\end{gather}
%
%
\begin{gather}
\lambda_r({r},0)=1, \quad \text{and} \quad \lambda_\theta({r},0)=1.
\end{gather}

As in the case of the outer constraint, we study the influence of the thickness (radius) and the yield stress of the constraining material on the percentage increase in length. We choose two different radii of the constraining material: $r_b=0.1$ and $r_b=0.5$. Fig.~\ref{fig:incons} shows the percentage increase in length for $r_b=0.1$ in (a) and (c), and for $r_b=0.5$ in (b) and (d). In (a) and (b), yield stress value in Region B is the same as that of Region C, while in (c) and (d) it is 10 times higher in Region B than in Region C. 

We observe that an increase in $r_b$ decreases the percentage increase in length because Region B undergoes a lower volumetric expansion compared to Region C. The continuity of displacements across the interface of Regions B and C imposes a constraint on the extent to which the silicon annulus (Region C) can expand in the axial direction. The greater the radius of the Region B, the higher is the constraining influence. Most interestingly, the percentage increase in length shows a significant reduction when the yield stress in Region B is 10 times higher than that in Region C \emph{and} the radius of Region B is also comparatively high (see panel (d)). To isolate and understand the influence of this increased yield stress, we compare (b) and (d) since the radius of Region B is the same in both. Considering a particular lithiation rate, ${J}_0=1$, for instance, we observe that in panel (d), there is a distinct transition in the percentage increase of length at SOC $\approx 350$ mAhg$^{-1}$and another one much further on, just before SOC $\approx 1500$ mAhg$^{-1}$. In contrast to this trend, in panel (b), there is only one transition in the percentage increase of length at SOC $\approx 100$ mAhg$^{-1}$. We probe the stress fields and the plastic stretches corresponding to these values of the SOC. We find that for the case of panel (d), in the region before the first transition (at SOC $\approx 350$ mAhg$^{-1}$), Region C starts to yield from the outside, and the yield region grows in size; Region B, however, does not yield at all. Beyond the transition point, the whole of Region C is in a state of yield, while Region B does not yield yet. Thus, after the first transition, axial deformations are delimited by the small elastic deformations of the inner constraining cylinder. At the same time, further volumetric expansion in the silicon of Region C is easily accommodated through plastic flow in the radial direction without the need to increase the length. This plastic flow significantly reduces the rate of length increases (shown by a drop in the gradient of the plot). For panel (b), however, the only transition occurring at SOC $\approx 100$ mAhg$^{-1}$ corresponds to a situation where both Regions B and C have yielded. Following this transition, further volumetric expansion due to increasing SOC values is easily accommodated by plastic flow in both Regions B and C which results in comparatively higher percentage increase in length. This influence of the higher yield stress is however hardly noticeable when the inner radius is very small as in panel (c) with $r_b=0.1$. \\

\section{Phase diagrams for the explicitly constrained cases} \label{sec:phase_diag}

\begin{figure}[ht!]
\centering
\begin{subfigure}[b]{0.5\textwidth}
\includegraphics[width=\textwidth]{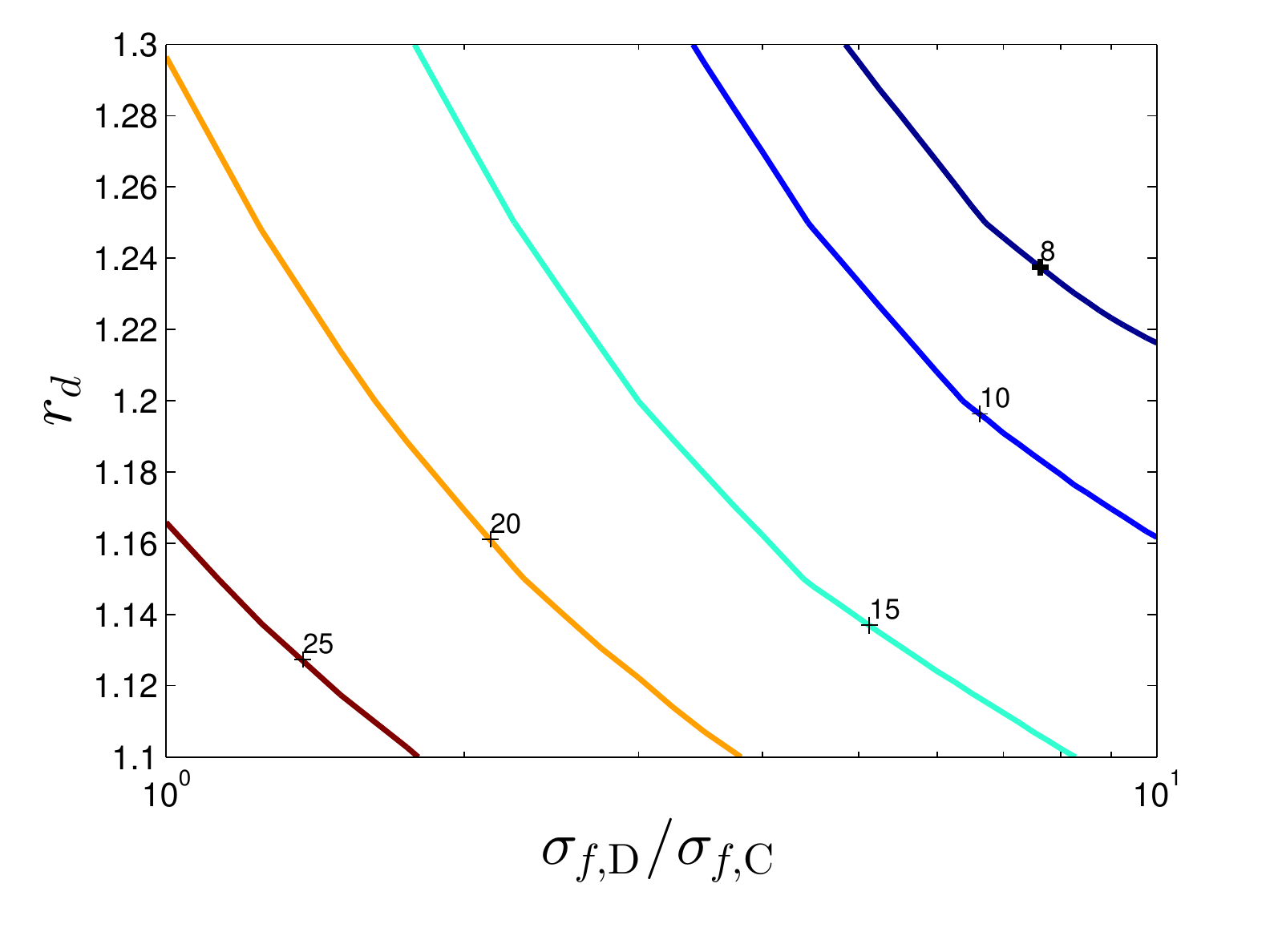}
\caption{}
\end{subfigure}%
\begin{subfigure}[b]{0.5\textwidth}
\includegraphics[width=\textwidth]{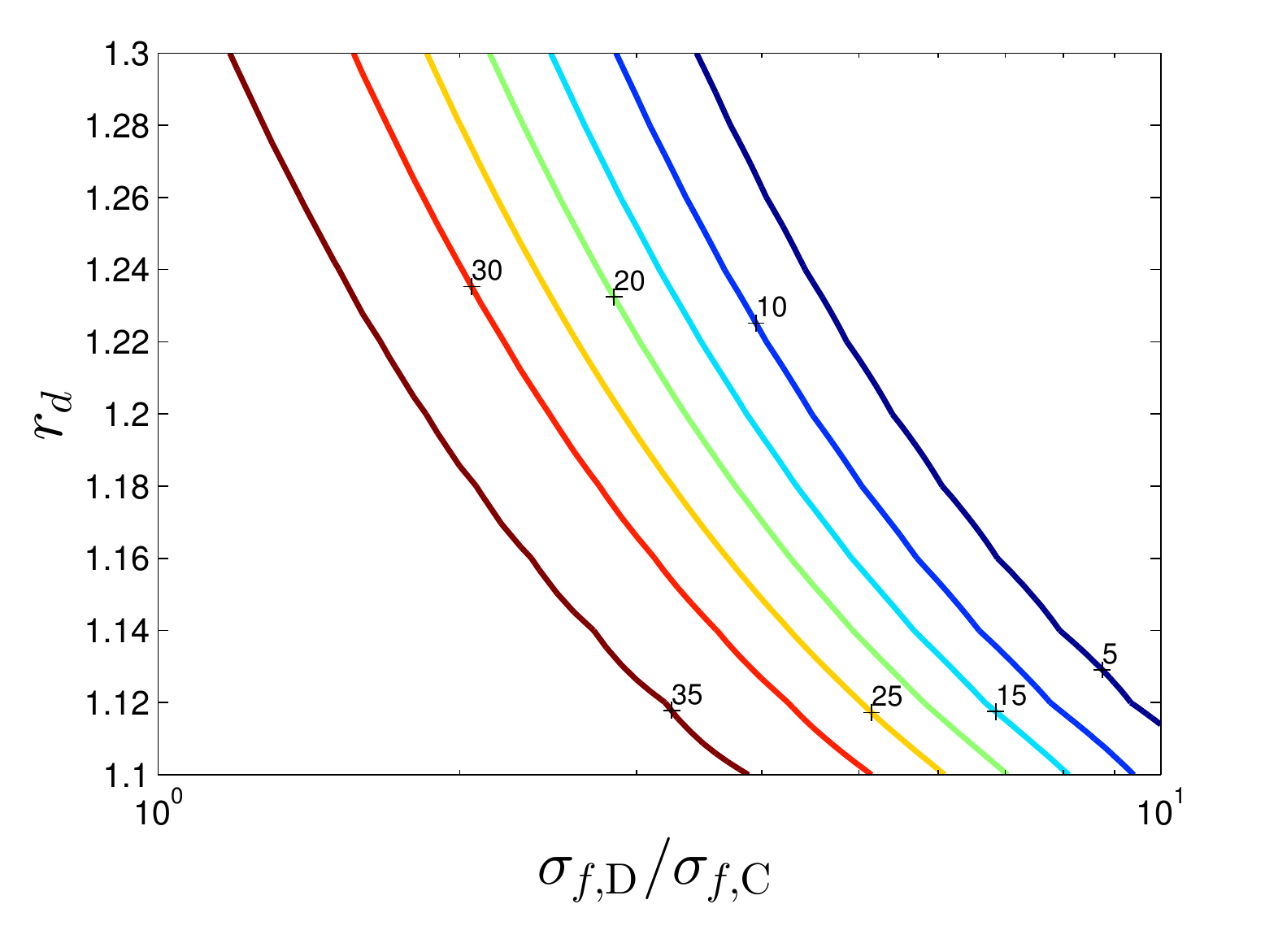}
\caption{}
\end{subfigure}
\caption{Contour plot of \% increase in length of a cylinder constrained from the outside. Different values of the ratio of the yield stress of the outer constraining annulus (Region D) and the Si cylinder (Region C) on abscissa; different values of radius of the constraining annulus (Region D) on the ordinate. (a) From simulation (b) From simplified model in Sec.~\ref{sec:sm2}.} \label{fig:outcons_phase_diag}
\end{figure}

Taking a cue from Figs.~\ref{fig:outcons}~(d) and \ref{fig:incons}~(d) which show that higher values of the yield stress and greater geometrical extent of the constraint result in smaller length increase, we study the combined influence of such factors through phase diagrams showing the percentage increase in length for the outer constraint as well as for the inner constraint case corresponding to different combinations of the thickness of the constraining material and the ratio of the yield stress of the constraining material to that of silicon. We do not present similar results for the simple annular cylinder because in that case we cannot define a ratio of the yield stresses. We first present the outer constraint case, and then the inner constraint case. 

\begin{figure}[ht!]
\centering
\begin{subfigure}[b]{0.5\textwidth}
\includegraphics[width=\textwidth]{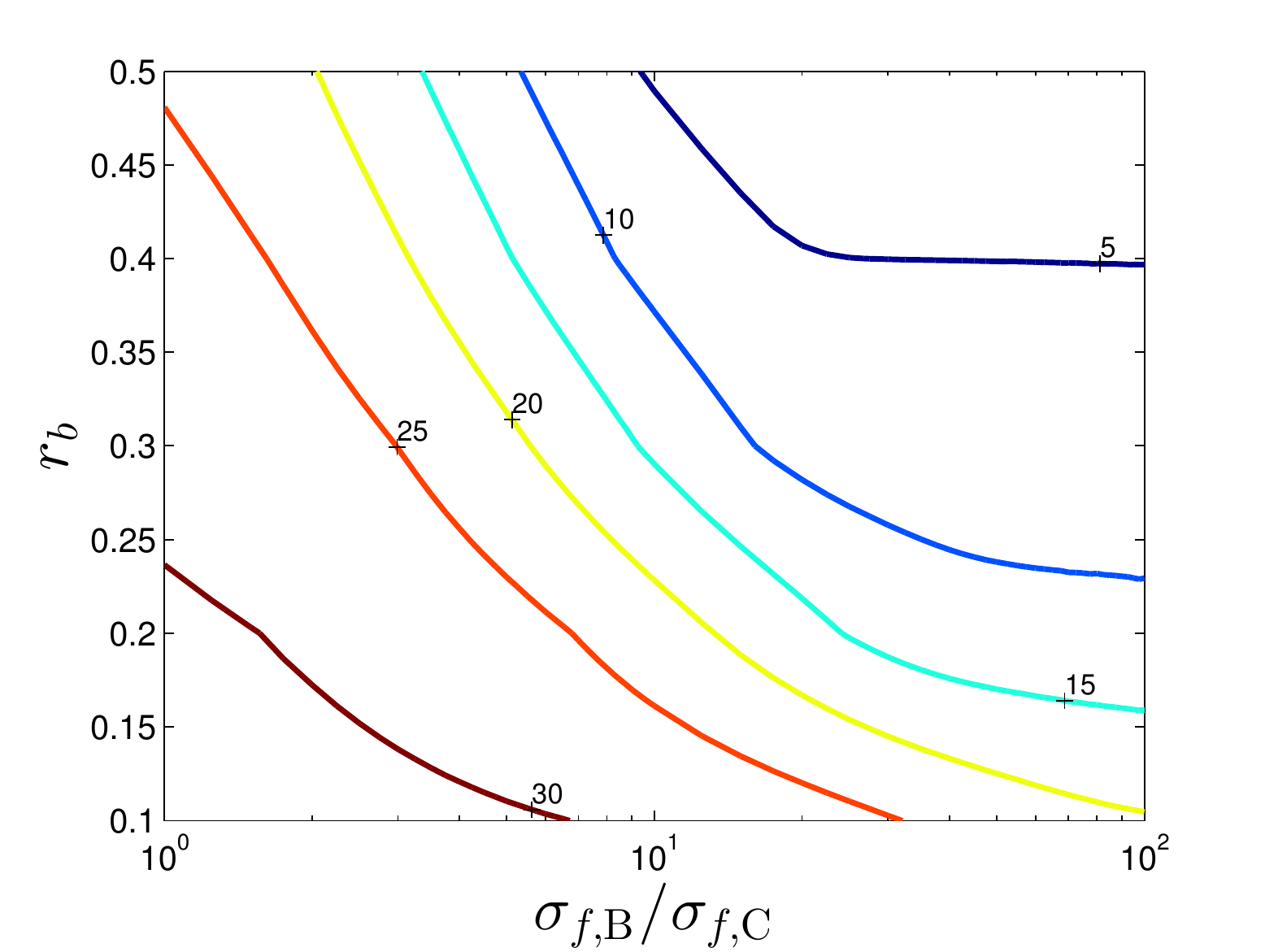}
\caption{}
\end{subfigure}%
\begin{subfigure}[b]{0.5\textwidth}
\includegraphics[width=\textwidth]{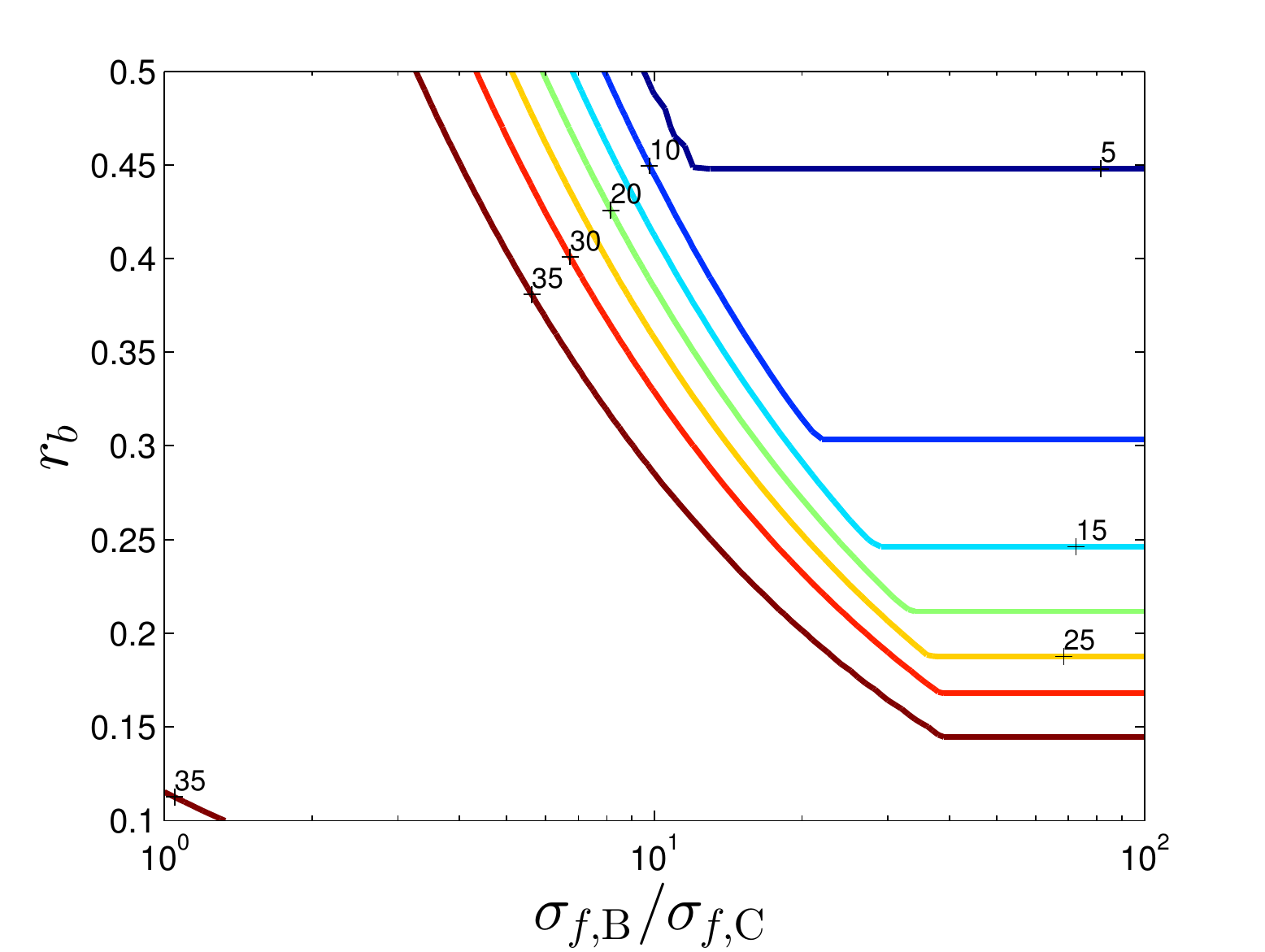}
\caption{}
\end{subfigure}
\caption{Contour plot of \% increase in length of a cylinder constrained from the inside. Different values of the ratio of the yield stress of the inner constraining cylinder and the outer annulus on abscissa; different values of radius of the constraining cylinder on the ordinate. (a) From simulation (b) From simplified model in Sec.~\ref{sec:sm2}.} \label{fig:incons_phase_diag}
\end{figure}

We map out in Figs.~\ref{fig:outcons_phase_diag} and \ref{fig:incons_phase_diag} a phase diagram of such percentage increases in length corresponding to various combinations of $r_d$ or $r_b$ and the ratio of the yield stress of the constraining material to that of silicon. We observe that the percentage increase in length is large for low values of the yield stress ratio and small geometrical extent of the constraint. Additionally, a decrease in the area of constraining region must be accompanied by an increase in the yield stress ratio in order to maintain the same percentage increase in length. In the case of the inner constraint, this finding is an important one for practical battery design because decreasing the inner radius allows decreasing the outer radius of the silicon annulus which is desirable from the perspective of reducing the overall volume. 

\begin{figure}[ht!]
\centering
\includegraphics[width=0.6\textwidth]{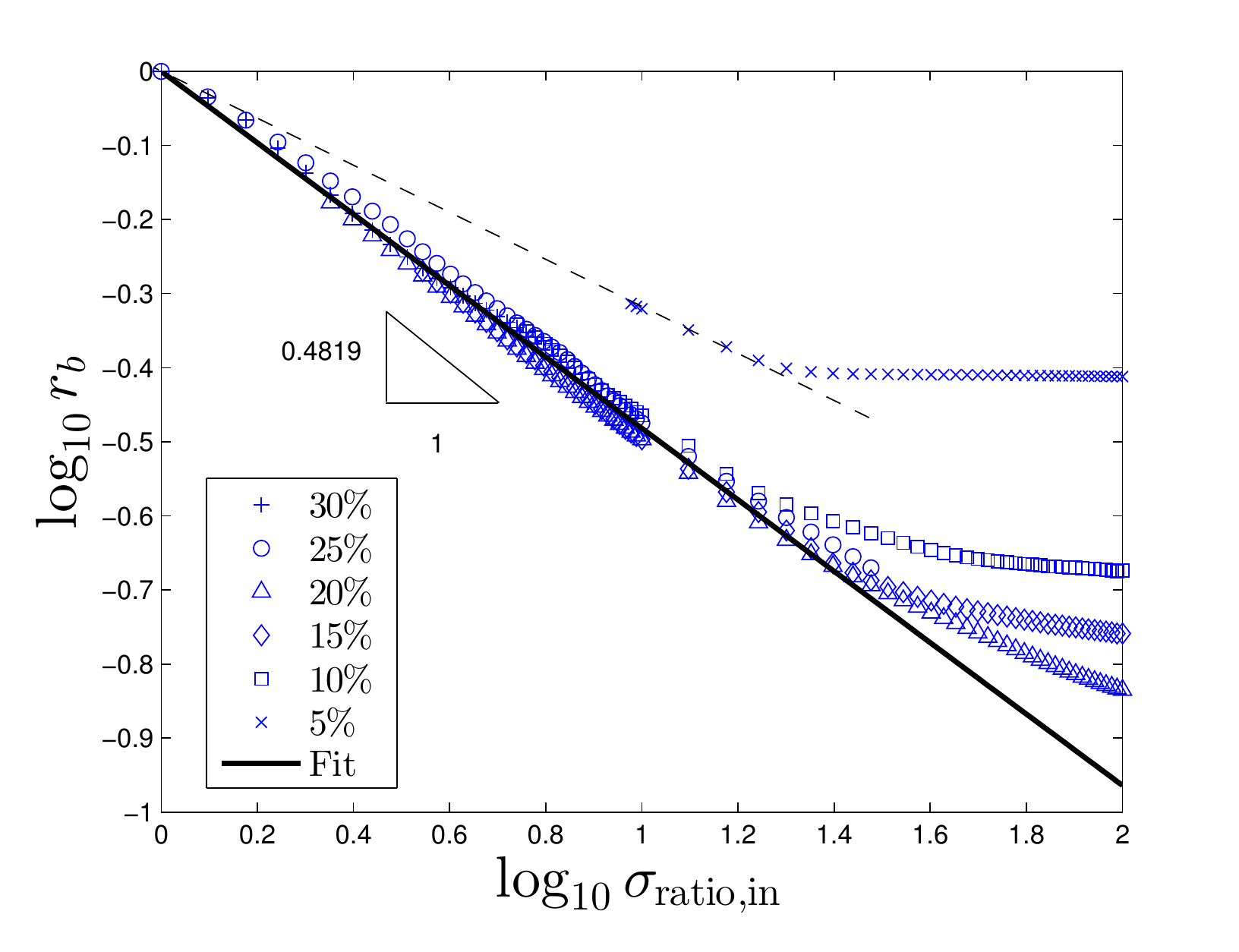}
\caption{Collapse of phase-diagram contours from Fig.~\ref{fig:incons_phase_diag}(a).} \label{fig:collapse_incons}
\end{figure}

When plotted on a log-log scale (with the intercepts along $\log_{10} r_b$ removed), the various contours of Fig.~\ref{fig:incons_phase_diag} are found to collapse, at least up to a yield stress ratio of around $1$, on to a straight line with slope approximately equal to $-0.5$. This is clearly seen in Fig.~\ref{fig:collapse_incons}. 


To interpret the phase diagram presented in Fig.~\ref{fig:incons_phase_diag}, we consider some simplified models which we describe now.

\section{First simplified model for the explicitly constrained cases} \label{sec:sm1}
We posit that the condition of zero net force in the axial direction is achieved through a simple force balance over the entire cross-section of the ``composite" cylinder (in the current or deformed configuration) in such a way that a net tensile force in one region is balanced exactly by a compressive force in the other region. In this simplified picture, we assume no spatial variation of the axial stress in each of the regions (silicon and constraint), and that (after yield) the axial stresses in each region are simply equal to the corresponding yield stresses. Further, the evolution of growth (and, hence, the stresses) is not tracked; rather a relation between the radius and the length change is obtained in the final deformed configuration with the assumption of negligible contributions from elastic deformation. The percentage increase in volume is taken to be 155\% corresponding to 50\% SOC. 

\subsection{Outer constraint} \label{subsec:pd_outcons}
The fractional increase in volume of the silicon cylinder (Region C) is given by
\begin{align}
\frac{(L+\Delta L) \pi (r_c + \Delta r_c)^2 - L \pi r_c^2}{L \pi r_c^2} &= 1.55, \\
{\rm{or,}} \quad 2 \Delta r_c + (\Delta r_c)^2 &= \frac{1.55-\kappa}{1+\kappa}, \label{eq:volC}
\end{align}
where $\kappa=\Delta L/L$, and where we have used $r_c=1$. Region D is assumed to undergo negligible volume change; thus
\begin{align}
(L+\Delta L) \pi [ (r_d + \Delta r_d)^2 - (r_c + \Delta r_c)^2 ] &= L \pi (r_d^2 - r_c^2), \\
{\rm{or,}} \quad  (r_d + \Delta r_d)^2 - (r_c + \Delta r_c)^2 &= \frac{r_d^2 - 1}{1+\kappa}. \label{eq:volD}
\end{align}
The axial force balance then gives
\begin{align}
\sigma_{\rm{C}} \pi (r_c + \Delta r_c)^2 &= \sigma_{\rm{D}} \pi [(r_d + \Delta r_d)^2 - (r_c + \Delta r_c)^2], \nonumber \\
{\rm{or,}} \quad r_d^2 &= 1 + \frac{2.55}{\sigma_{\rm{ratio,out}}}, \label{eq:simple_rd}
\end{align}
where $\sigma_{\rm{ratio,out}}=\sigma_{\rm{D}}/\sigma_{\rm{C}}$ with $\sigma_{\rm{C}}$ and $\sigma_{\rm{D}}$ being the constant (magnitudes of the) stresses in Regions C and D, and where we have used both (\ref{eq:volC}) and (\ref{eq:volD}). 

\begin{figure}
\centering
\includegraphics[width=0.5\textwidth]{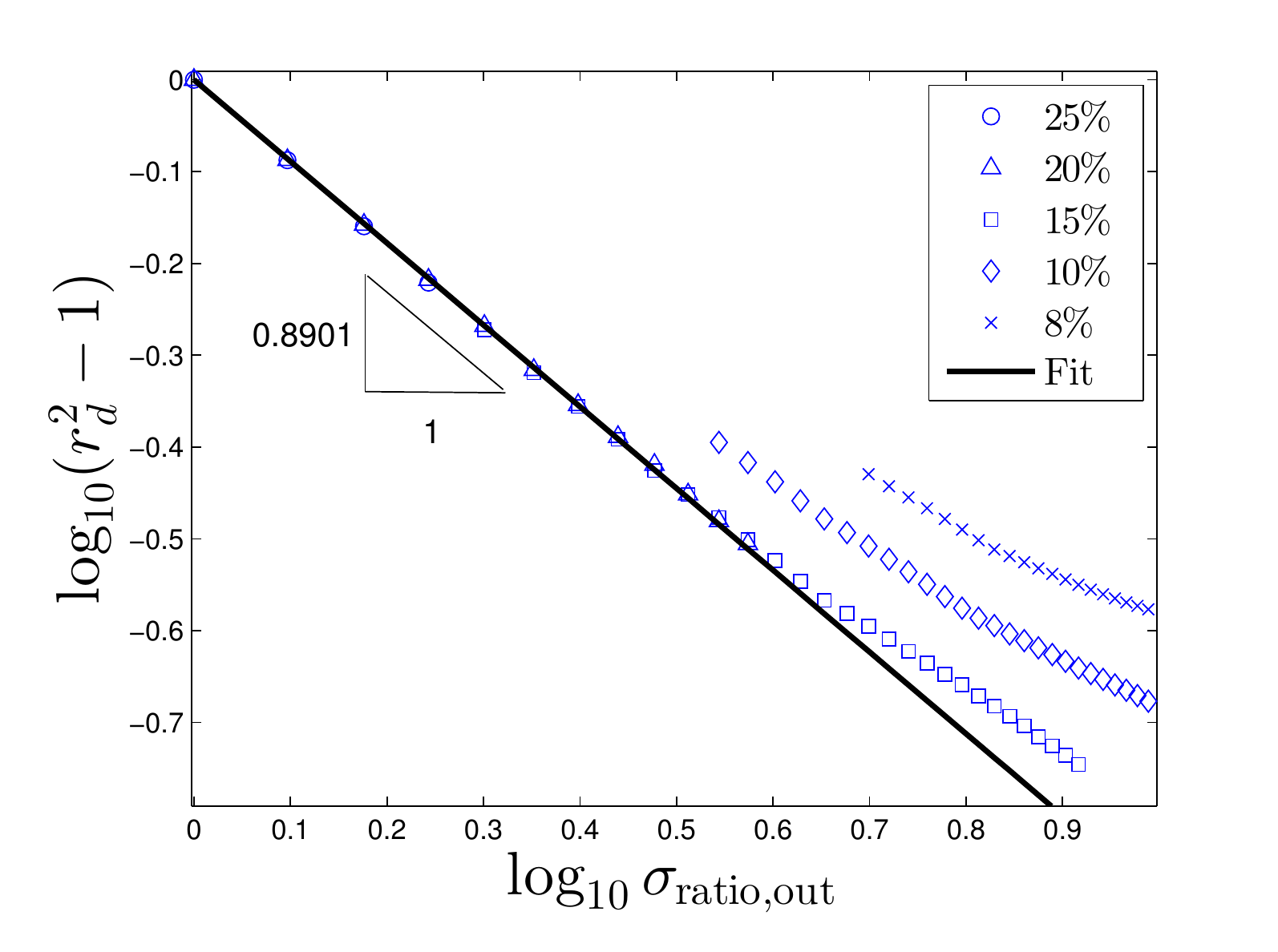}
\caption{Collapse of phase-diagram contours from Fig.~\ref{fig:outcons_phase_diag}(a).} \label{fig:collapse_outcons}
\end{figure}

To specify $\sigma_{\rm{C}}$ and $\sigma_{\rm{D}}$ we assume that both Regions C and D have yielded plastically by the time 50\% SOC is reached. For such a situation, both $\sigma_{\rm{C}}$ and $\sigma_{\rm{D}}$ may be expected to be at their respective yield stress values. Guided by Eq.~(\ref{eq:simple_rd}), a log-log plot of the raw simulation data shows a collapse (barring two outliers) of the $\log_{10}(r_d^2-1)$ vs $\log_{10}\sigma_{\rm{ratio,out}}$ points on to a straight line (at least up to $\log_{10} \sigma_{\rm{ratio,out}}=0.5$) with a slope of $-0.89$ as seen in Fig.~\ref{fig:collapse_outcons}. This is in good agreement with the scaling relationship from Eq.~(\ref{eq:simple_rd}) of $r_d^2-1 \sim \sigma_{\rm{ratio,out}}^{-1}$. 

\subsection{Inner constraint} \label{subsec:pd_incons}

The fractional increase in volume of the silicon annulus (Region C) is given by
\begin{align}
\frac{(L+\Delta L) \pi \left[ (r_c + \Delta r_c)^2 - r_b^2 \right] - L\pi (r_c^2 - r_b^2)}{L \pi (r_c^2 - r_b^2)} &= 1.55, \\
{\rm{or,}} \quad (\Delta r_c)^2 + 2 r_c \Delta r_c &= \frac{1.55 - \kappa}{1+\kappa},
\end{align}
where we have used $r_c^2 - r_b^2 = 1$. With the assumption that the radius of the constraint remains practically unchanged (based on the fact that its volumetric expansion is taken to be negligibly small compared to the silicon), the force balance gives
\begin{align}
\sigma_{\rm{B}} \pi r_b^2 &= \sigma_{\rm{C}} \pi \left[ (r_c + \Delta r_c)^2 - r_b^2  \right], \\
{\rm{or,}} \quad r_b^2 &= \frac{2.55}{\sigma_{\rm{ratio,in}}(1+\kappa)}, \label{eq:simple_rb}
\end{align}
where $\sigma_{\rm{B}}$ and $\sigma_{\rm{C}}$ are the constant (magnitudes of the) axial stresses in Regions B and C respectively, and $\sigma_{\rm{ratio,in}}=\sigma_{\rm{B}}/\sigma_{\rm{C}}$. To specify $\sigma_{\rm{B}}$ and $\sigma_{\rm{C}}$ we need to consider different regimes of mechanical behaviour of Regions B and C.

We expect that when the yield stress of Region B is not too high, then both Regions B and C have yielded plastically by the time 50\% SOC is reached. Both $\sigma_{\rm{B}}$ and $\sigma_{\rm{C}}$ may then be expected to be at their respective yield stress values. Therefore, we take $\sigma_{\rm{ratio,in}}=\sigma_{f,{\rm{B}}}/\sigma_{f,{\rm{C}}}$ which approximates the left-most section of each contour in Fig.~\ref{fig:incons_phase_diag}. For high enough yield stresses in Region B, it might not yield even though Region C does. For such a situation, $\sigma_{\rm{B}}=E \kappa$ ($E$ being the modulus of elasticity of the constraining material which is considered to be the same as that of Si, $\kappa$ being the fractional increase in length), and this is a good approximation for relatively low values of $\kappa$. For Region C in a state of yield, $\sigma_{\rm{C}} = \sigma_{f,\rm{C}}$. Using this in Eq.~\ref{eq:simple_rb} gives horizontal lines which approximate the right-most part of each contour in Fig.~\ref{fig:incons_phase_diag}. However the model is not quantitatively accurate in predicting axial length increases because we have assumed that the axial stresses are equal to the yield stress, whereas the charging rates are such that in the simulations, the axial stresses in both Regions C and D increase much beyond the respective yield stress values. 
Through Eq.~(\ref{eq:simple_rb}), this simple model has predicted the correct scaling ratio, $r_b \sim \sigma_{\rm{ratio,in}}^{-1/2}$, observed in the collapsed plots of Fig.~\ref{fig:collapse_incons}.

\section{Second simplified model} \label{sec:sm2}

\begin{figure}[ht!]
\centering
\includegraphics[width=0.6\textwidth]{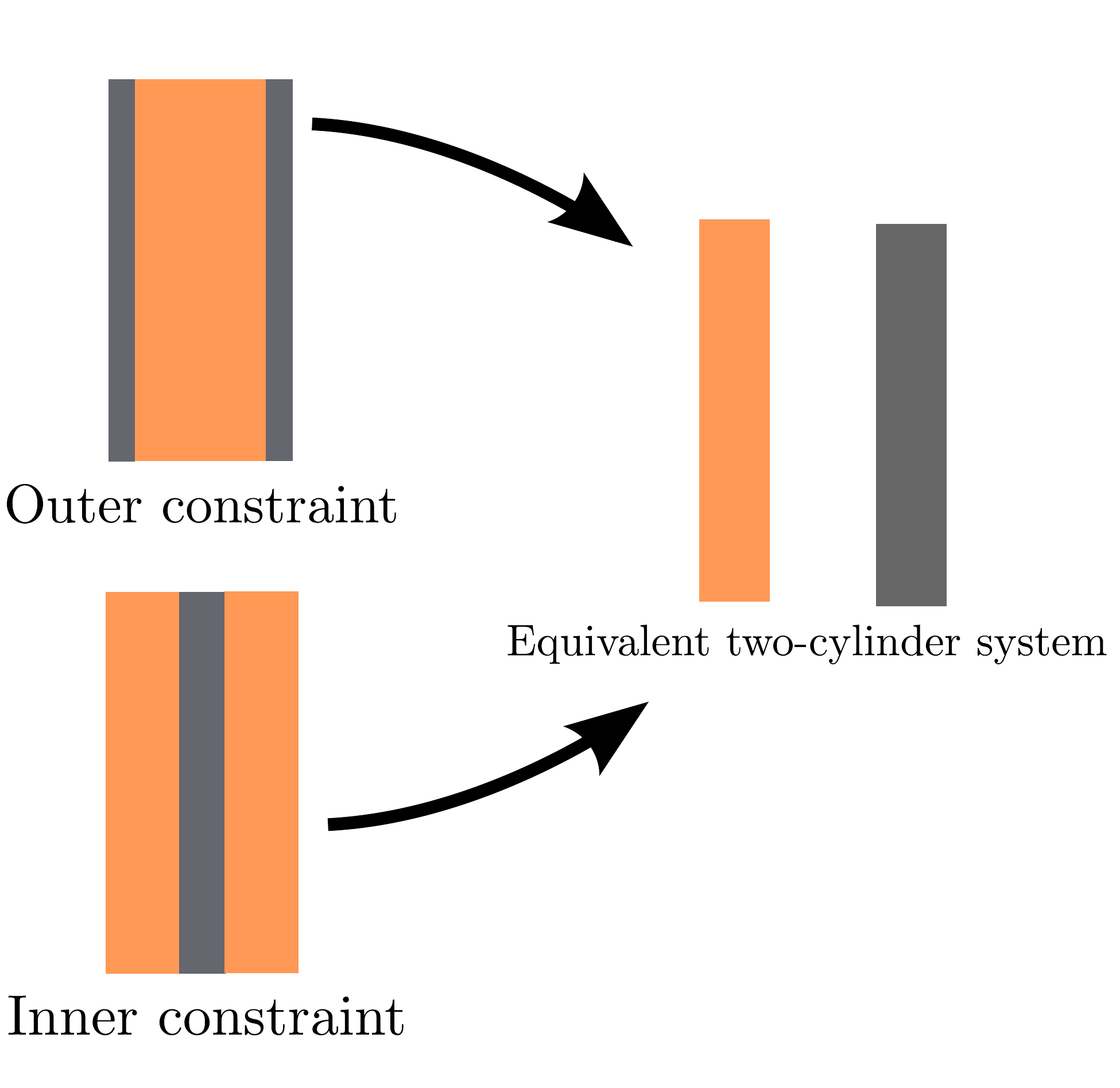}
\caption{Equivalent two-cylinder system for both outer and inner constraint cases.} \label{fig:sm2}
\end{figure}

We develop a second simplified model to capture the evolution of the stresses and the plastic stretches for the case of the annular silicon cylinder with an outer or an inner constraint. Here, we model the annular cylinder-constraint system by an equivalent two cylinder model, as shown in Fig.~\ref{fig:sm2}. Thus, Regions B, C, and D are each modelled by solid circular cylinders. Guided by the phase-diagrams and the first simplified model, we consider two regimes. In the first one, we consider that both the cylinders are flowing plastically, whereas in the second one, we consider that while the cylinder made of silicon has yielded, the cylinder made of the constraining material has not (i.e. it stays elastic). Note that the first (plastic-plastic regime) is applicable to the inner constraint case as well as the outer constraint one. The second regime, however, is applicable only to the inner constraint case; in other words, the system with the outer constraint is assumed never to operate within the elastic-plastic regime. Further, we also assume that the lithium concentration in the silicon cylinder is not affected by the stresses, and is, therefore, known independent of any mechanical considerations. In what follows, we consider the two regimes (plastic-plastic and elastic-plastic) separately. Within the first regime, we discuss the outer and the inner constraint cases, and within the second regime, we discuss only the inner constraint case.

\subsection{Plastic constraint-Plastic Si regime}
This regime is applicable for both the outer and the inner constraint cases. We assume separable solutions in the form of
\begin{subequations} \label{eq:sepsoln}
\begin{align}
u &= A(t) r, \\
w &= B(t) r.
\end{align}
\end{subequations}
Further assuming small strains, we have
\begin{subequations} \label{eq:small-strain}
\begin{align}
1+\frac{\partial u}{ \partial r} &= \lambda_r (J^c)^{1/3} \\
1+\frac{u}{r} &= \lambda_\theta (J^c)^{1/3} \\
1+\frac{\partial w}{\partial z} &= \lambda_z (J^c)^{1/3}.
\end{align}
\end{subequations}
Substituting Eq.~(\ref{eq:sepsoln}) in Eq.~(\ref{eq:small-strain}a) and (b), we have
\begin{align}
\lambda_r = \lambda_\theta =: \lambda, \\
1+A = \lambda (J^c)^{1/3},
\end{align}
and from Eq.~(\ref{eq:small-strain}c), we have
\begin{gather}
1+B = \frac{1}{\lambda^2} (J^c)^{1/3},
\end{gather}
where we have used $\lambda_z = 1/(\lambda_r \lambda_\theta)$.

Again using the smallness of the strains, we have for the Piola-Kirchhoff stresses:
\begin{subequations} \label{eq:redPK}
\begin{align}
\sigma_r^0 &= \frac{J^c}{1+A}\frac{E}{(1+\nu)(1-2\nu)} \left[ (1-\nu)E_r^e + \nu(E_\theta^e + E_z^e)  \right], \\
\sigma_\theta^0 &= \frac{J^c}{1+A} \frac {E}{(1+\nu)(1-2\nu)} \left[ (1-\nu)E_\theta^e + \nu(E_z^e + E_r^e)  \right], \\
\sigma_z^0 &= \frac{J^c}{1+B} \frac {E}{(1+\nu)(1-2\nu)} \left[ (1-\nu)E_z^e + \nu(E_r^e + E_\theta^e)  \right].
\end{align}
\end{subequations}
Similarly for the Cauchy stresses, we have:
\begin{subequations} \label{eq:redCauchy}
\begin{align}
\sigma_r &= \frac{E}{(1+\nu)(1-2\nu)} \left[ (1-\nu)E_r^e + \nu(E_\theta^e + E_z^e)  \right], \\
\sigma_\theta &=  \frac {E}{(1+\nu)(1-2\nu)} \left[ (1-\nu)E_\theta^e + \nu(E_z^e + E_r^e)  \right], \\
\sigma_z &= \frac {E}{(1+\nu)(1-2\nu)} \left[ (1-\nu)E_z^e + \nu(E_r^e + E_\theta^e)  \right].
\end{align}
\end{subequations}
Comparing Eq.~(\ref{eq:redPK}) and Eq.~(\ref{eq:redCauchy}), we have:
\begin{gather} \label{eq:PK-Cauchy}
\sigma_r^0 = \frac{J^c}{1+A} \sigma_r, \quad \sigma_\theta^0 = \frac{J^c}{1+A}\sigma_\theta, \quad \text{and} \quad \sigma_z^0 = \frac{J^c}{1+B} \sigma_z. 
\end{gather}
The plastic stretch evolution equations in the radial and the hoop directions are, respectively,
\begin{subequations} \label{eq:plastic-stretch}
\begin{align}
\frac{\dot{\lambda}_r}{\lambda_r} = \sqrt{\frac{3}{2}} {\rm{Pf}} \left( \frac{\sigma_{\rm{eff}}}{\sigma_{\rm{f}}} - 1 \right)^m H\left( \frac{\sigma_{\rm{eff}}}{\sigma_{\rm{f}}} - 1 \right) \frac{\tau_r}{\| \bm{\tau} \|}, \\
\frac{\dot{\lambda}_\theta}{\lambda_\theta} = \sqrt{\frac{3}{2}} {\rm{Pf}} \left( \frac{\sigma_{\rm{eff}}}{\sigma_{\rm{f}}} - 1 \right)^m H\left( \frac{\sigma_{\rm{eff}}}{\sigma_{\rm{f}}} - 1 \right) \frac{\tau_\theta}{\| \bm{\tau} \|}, 
\end{align}
\end{subequations}
where ${\rm{Pf}}=\dot{d}_0R_0^2/D_0$. Since $\lambda_r = \lambda_\theta = \lambda$ and they have the same initial condition of $\lambda=1$, Eq.~(\ref{eq:plastic-stretch}) implies that $\tau_r = \tau_\theta$. That is, $\sigma_r = \sigma_\theta$. Therefore, from Eq.~(\ref{eq:PK-Cauchy}), we get $\sigma_r^0 = \sigma_\theta^0$. Using this in the mechanical equilibrium equation, we have:
\begin{gather}
\frac{\partial \sigma_r^0}{\partial r} = 0,
\end{gather}
which implies that $\sigma_r^0$ is spatially invariant. If we use the traction-free boundary condition then we have $\sigma_r^0 = \sigma_\theta^0 = 0$. And, this, in turn, implies from Eq.~(\ref{eq:PK-Cauchy}) that $\sigma_r = \sigma_\theta = 0$. In this situation, we have the following:
\begin{gather}
\tau_r = -\frac{1}{3} \sigma_z, \quad \tau_\theta = -\frac{1}{3} \sigma_z, \quad \text{and} \quad \tau_z = \frac{2}{3} \sigma_z.
\end{gather}
Further,
\begin{align}
\sigma_{\rm{eff}} &= \sqrt{\frac{3}{2}} \sqrt{\tau_r^2 + \tau_\theta^2 + \tau_z^2} = |\sigma_z|, \\
\|\bm{\tau}\| &= \sqrt{\tau_r^2 + \tau_\theta^2 + \tau_z^2} = \sqrt{\frac{2}{3}} |\sigma_z|.
\end{align}
Then the (only independent) platic stretch evolution equation becomes
\begin{gather}
\frac{\dot{\lambda}}{\lambda} = {\rm{Pf}} \left( \frac{|\sigma_z|}{\sigma_{\rm{f}}} - 1\right)^m H\left( \frac{|\sigma_z|}{\sigma_{\rm{f}}} - 1\right) {\rm{sgn}}(\sigma_z).
\end{gather}
Since we are assuming that both regions have yielded, the Heaviside function may be dropped. 

In the following, we use the subscript ``con" to denote Region B or D. It is also important to note that in the constraint cylinder there is no lithium influx; therefore, in these regions we have $J^c=1$. Thus for the two cylinder system we have the following equations:

\textbf{Region B or Region D}
\begin{align}
\frac{\dot{\lambda}_{\rm{con}}}{\lambda_{\rm{con}}} &= -\frac{1}{2} {\rm{Pf}} \left( \frac{|\sigma_{z,{\rm{con}}}|}{\sigma_{f,{\rm{con}}}} - 1 \right)^m {\rm{sgn}}(\sigma_{z,{\rm{con}}}), \label{eq:con-first}\\
1 + B_{\rm{con}} &= \frac{1}{\lambda_{\rm{con}}^2}. \label{eq:con-second}
\end{align}

\textbf{Region C}
\begin{align}
\frac{\dot{\lambda}_{\rm{C}}}{\lambda_{\rm{C}}} &= -\frac{1}{2} {\rm{Pf}} \left( \frac{|\sigma_{z,{\rm{C}}}|}{\sigma_{f,{\rm{C}}}} - 1 \right)^m {\rm{sgn}}(\sigma_{z,{\rm{C}}}), \label{eq:C-first}\\
1+B_{\rm{C}} &= \frac{1}{\lambda_{\rm{C}}^2} (J^c)^{1/3}. \label{eq:C-second}
\end{align}
In this system, we have six unknowns: $\lambda_{\rm{con}}$, $\lambda_{\rm{C}}$, $\sigma_{z,{\rm{con}}}$, $\sigma_{z,{\rm{C}}}$, $B_{\rm{con}}$, and $B_{\rm{C}}$, and four equations. We, therefore, need two more equations to close the system. The first of these equations is given by the condition that the percentage increase in length of both cylinders should be the same; thus:
\begin{gather}
B_{\rm{con}} = B_{\rm{C}}. \label{eq:equal-length}
\end{gather}
The final equation is given by the condition of force-balance in the axial direction of the two-cylinder system which models the fact that the net force in the axial direction in the original annular-constraint system is zero. Thus, in the reference configuration we have
\begin{align}
a_{\rm{con}}^0 |\sigma_{z,{\rm{con}}}^0| = a_{\rm{C}}^0 |\sigma_{z,{\rm{C}}}^0|, \nonumber \\
\Rightarrow |\sigma_{z,{con}}| = a J^c |\sigma_{z,{\rm{C}}}| \label{eq:force-balance}.
\end{align}
Here, $a_{\rm{con}}^0$ is either $a_{\rm{D}^0}=r_d^2-1$ or $a_{\rm{B}}^0 = \pi r_b^2$, and $a_{\rm{C}}^0 = \pi$ are the areas of the two cylinders in the reference configuration, and $a=a_{\rm{C}}^0/a_{\rm{con}}^0$ is the ratio of the two reference areas.

Comparing Eq.~(\ref{eq:con-second}) and Eq.~(\ref{eq:C-second}), we have
\begin{gather}
\frac{1}{\lambda_{\rm{con}}^2} = \frac{1}{\lambda_{\rm{C}}^2} (J^c)^{1/3}.
\end{gather}
Taking logarithms and differentiating with respect to time, we obtain
\begin{align}
\frac{\dot{\lambda}_{\rm{con}}}{\lambda_{\rm{con}}} &= \frac{\dot{\lambda}_{\rm{C}}}{\lambda_{\rm{C}}} - \frac{1}{6} \frac{\dot{J}^c}{J^c}, \nonumber \\
\Rightarrow -\frac{1}{2} {\rm{Pf}} \left( \frac{|\sigma_{z,{\rm{con}}}|}{\sigma_{f,{\rm{con}}}} - 1 \right)^m {\rm{sgn}}(\sigma_{z,{\rm{con}}}) &= -\frac{1}{2} {\rm{Pf}} \left( \frac{|\sigma_{z,{\rm{C}}}|}{\sigma_{f,{\rm{C}}}} - 1 \right)^m {\rm{sgn}}(\sigma_{z,{\rm{C}}}) - \frac{1}{6} \frac{\dot{J}^c}{J^c}. \label{eq:iter-intm}
\end{align}
If we let $|\sigma_{z,{\rm{C}}}|=\sigma_{f,{\rm{C}}} \Sigma$, then using  Eq.~(\ref{eq:force-balance}) in Eq.~(\ref{eq:iter-intm}) leads to
\begin{gather}
\left( \frac{aJ^c}{\sigma_{f,{\rm{ratio}}}}\Sigma - 1 \right)^m {\rm{sgn}}(\sigma_{z,{\rm{B}}}) = \left( \Sigma - 1 \right)^m {\rm{sgn}}(\sigma_{z,{\rm{C}}}) + \frac{1}{3{\rm{Pf}}}\frac{\dot{J}^c}{J^c}, \label{eq:iter-final}
\end{gather}
where $\sigma_{f,{\rm{ratio}}}=\sigma_{f,{\rm{con}}}/\sigma_{f,{\rm{C}}}$. The sign functions may be resolved from a physical understanding of the situation. The length increase of the Region C cylinder is due to lithiation-induced growth, and under the condition of zero net axial force this growth is constrained; therefore, the axial stresses are expected to be compressive, implying ${\rm{sgn}(\sigma_{z,{\rm{C}}})}=-1$. The length increase in the Region D or Region B cylinder is to due to stretching so that Eq.~(\ref{eq:equal-length}) is satisfied; therefore, ${\rm{sgn}}(\sigma_{z,{\rm{con}}})=1$. Eq.~(\ref{eq:iter-final}) is a non-linear algebraic equation in $\Sigma$, and is solved iteratively for each time step given the values of $J^c$ and $\dot{J}^c$. This value of $\Sigma$ is then used to advance the plastic stretch using Eq.~(\ref{eq:C-first}), following which the increase in length, $B_{\rm{C}}$, of the Region C cylinder (and, hence, of the two-cylinder system) is found from Eq.~(\ref{eq:C-second}). 

Note from Eq.~(\ref{eq:iter-final}) that $\Sigma$ depends on $\sigma_{f,{\rm{ratio}}}$ and $r_b$ only through $a/\sigma_{f,{\rm{ratio}}}$. Thus this model also predicts that contours of constant axial extension should satisfy the scaling law $r_b \tilde \sigma_{f,{\rm{ratio}}}^{-1/2}$.  

\subsection{Elastic constraint-Plastic Si regime}
We consider the Region B cylinder to be in the elastic regime, and the Region C cylinder to have yielded. Again, we start by assuming separable solutions of the form Eq.~(\ref{eq:sepsoln}). But since $\lambda_r = \lambda_\theta = \lambda_z = 1$ in Region B, we cannot take $E_z^e$ to be vanishingly small to obtain relations akin to Eq.~(\ref{eq:small-strain}) because that would result in $B=0$ (note that $J^c=1$ in Region B). However, we observe that
\begin{align}
u = Ar \Rightarrow E_r^e = E_\theta^e.
\end{align}
Then, the expressions of the Piola-Kirchhoff stresses in the radial and the hoop directions, respectively, become:
\begin{subequations} \label{eq:elPK}
\begin{align}
\sigma_r^0 = \frac{E}{(1+\nu)(1-2\nu)} \left[ E_r^e + \nu E_z^e  \right]\frac{2E_r^e+1}{1+A}, \\
\sigma_\theta^0 = \frac{E}{(1+\nu)(1-2\nu)} \left[ E_\theta^e + \nu E_z^e  \right]\frac{2E_\theta^e+1}{1+A},
\end{align}
\end{subequations}
from which we infer $\sigma_r^0 = \sigma_\theta^0$. From the mechanical equilibrium equation, this equality, in turn, implies
\begin{gather}
\frac{\partial \sigma_r^0}{\partial r} = 0.
\end{gather}
Again using the traction-free boundary condition, this gives us $\sigma_r^0 = 0$. Therefore, $\sigma_\theta^0 = 0$, too. Then, from Eq.~(\ref{eq:elPK}), we have
\begin{gather}
E_r^e = E_\theta^e = -\nu E_z^e. 
\end{gather}
Using this in the expression for $\sigma_z^0$ and simplifying, we obtain
\begin{gather}
\sigma_z^0 = E E_z^e \frac{2E_z^e+1}{1+B}. \label{eq:elPKz}
\end{gather}
Similarly, we obtain for the Cauchy stress in the axial direction:
\begin{gather}
\sigma_z = E E_z^e \frac{2E_z^e + 1}{(1+A)^2(1+B)}. \label{eq:elCauchyz}
\end{gather}
Comparing Eq.~(\ref{eq:elPKz}) and Eq.~(\ref{eq:elCauchyz}), we obtain the relation:
\begin{gather}
\sigma_z^0 = (1+A)^2 \sigma_z \label{eq:elPK-Cauchy}
\end{gather}
We now have the following equations for the two-cylinder system:

\textbf{Region B cylinder:}
\begin{align}
2E_z^e + 1 &= (1+B)^2, \label{eq:elB-first}\\
2E_r^e + 1 &= (1+A_{\rm{B}})^2, \label{eq:elB-second}\\
E_r^e &= -\nu E_z^e, \label{eq:elB-third}\\
\sigma_{z,{\rm{B}}} &= E E_z^e \frac{2E_z^e+1}{(1+A)^2(1+B)}. \label{eq:elB-fourth} 
\end{align}

\textbf{Region C cylinder:}
\begin{align}
1+B &= \frac{1}{\lambda^2}(J^c)^{1/3}, \label{eq:elC-first} \\
1+A_{\rm{C}} &= \lambda (J^c)^{1/3}, \label{eq:elC-second} \\
\frac{\dot{\lambda}}{\lambda} &= - \frac{1}{2} {\rm{Pf}} \left( \frac{|\sigma_{z,{\rm{C}}}|}{\sigma_{f,{\rm{C}}}} - 1  \right)^m H\left( \frac{|\sigma_{z,{\rm{C}}}|}{\sigma_{f,{\rm{C}}}} - 1  \right) {\rm{sgn}} (\sigma_{z,{\rm{C}}}). \label{eq:elC-third}
\end{align}
Note that we have retained the Heaviside function here. In this system, we have eight unknowns: $E_z^e$, $B$, $E_r^e$, $A_{\rm{B}}$, $A_{\rm{C}}$, $\lambda$, $\sigma_{z,{\rm{B}}}$, and $\sigma_{z,{\rm{C}}}$, and seven equations. Therefore, we need another equation to close the system. This is provided again by the condition of force balance just as in the plastic-plastic regime:
\begin{align}
|\sigma_{z,{\rm{B}}}^0| &= a |\sigma_{z,{\rm{C}}}^0|, \nonumber \\
\Rightarrow  |\sigma_{z,{\rm{B}}}|(1+A_{\rm{B}})^2 &= a \frac{J^c}{1+B} |\sigma_{z,C}|. \label{eq:elforce-balance-intm}
\end{align}
Now, linearizing Eq.~(\ref{eq:elB-first}) in $B$, we obtain
\begin{gather}
E_z^e \approx B.
\end{gather}
Using this in Eq.~(\ref{eq:elB-fourth}), and linearizing again in $B$, we have
\begin{align}
\sigma_{z,{\rm{B}}} \approx E \frac{B}{(1+A_{\rm{B}})^2(1+B)}. 
\end{align}
Using this in Eq.~(\ref{eq:elforce-balance-intm}), we obtain
\begin{align}
|\sigma_{z,{\rm{C}}}| = \frac{B E}{a J^c}. \label{eq:elforce-balance}
\end{align}
If we let, as before, $|\sigma_{z,{\rm{C}}}| = \sigma_{f,{\rm{C}}}\Sigma$, then we have from Eq.~(\ref{eq:elforce-balance}) and Eq.~(\ref{eq:elC-first}) the following:
\begin{align}
\Sigma = \left( \frac{E}{\sigma_{f,{\rm{C}}}} \right) \frac{1}{aJ^c} \left[ \frac{(J^c)^{1/3}}{\lambda^2} - 1  \right]. \label{eq:elSigma}
\end{align}
Further, from Eq.~(\ref{eq:elC-third}), we have
\begin{align}
\frac{\dot{\lambda}}{\lambda} = -\frac{1}{2} {\rm{Pf}} \left( \Sigma - 1  \right)^m H\left( \Sigma - 1 \right) {\rm{sgn}}(\sigma_{z,{\rm{C}}}). \label{eq:ellambda}
\end{align}
At any time step, Eq.~(\ref{eq:elSigma}) is used to find the value of $\Sigma$ given $J^c$ and the value of $\lambda$ for that time step. This is then used in Eq.~(\ref{eq:ellambda}) to advance the plastic stretch to the next time step. The phase-diagrams from this second simplified model for the outer constraint and the inner constraint case are shown in Figs.~\ref{fig:outcons_phase_diag}~(b) and \ref{fig:incons_phase_diag}~(b), respectively. It can be seen that these phase diagrams show the same qualitative trends as those seen in the phase diagrams, Figs.~\ref{fig:outcons_phase_diag}~(a) and (b), obtained from the numerical simulations. 

\section{Conclusions} \label{sec:conclusions}

It is generally accepted that a cylindrical geometry may overcome the commonly faced problem of pulverization of anode particles made of silicon following extreme volume changes during cyclic charging and discharging. However, the volume expansion itself can still lead to problems due to the finite confines of a battery. In the presence of other particles or supporting structures, the volume expansion will be constrained, and can lead to the generation of stresses, leading - as we have investigated in our previous work \cite{2014IJSSJeevanAdvisors} - to the possibility of failure by buckling. Motivated by the need to mitigate these problems arising from constrained volume expansion, we look for ways to modify the cylindrical geometry in the simplest possible ways so that expansion, particularly in the axial direction, may be reduced. We first studied an annular geometry noting the differences in axial growth when lithiated from the inside and the outside at different flux rates. We found that the annular geometry does not provide much respite from the problem of axial growth compared to the cylindrical geometry. Next we investigated the possibilities of reducing the axial growth by constraining a solid cylinder from the outside, and an annular cylinder from the inside. The constraining material is taken to be such that its volumetric expansion is negligible compared to that of silicon. We found that in both cases axial growth can indeed by substantially reduced by a material whose yield stress is significantly higher than that of silicon. This reduction is further reinforced by increasing the thickness of the outer constraint or the radius of the inner constraint (maintaining the same volume of silicon in all cases). We presented phase diagrams for both the outer and the inner constraint cases, plotting axial growth for various combinations of the radius and the yield stress ratio of the constraint and silicon, to identify desirable operating zones. Since all our results are based on a fully-coupled mechano-chemical model that is amenable only to numerical solutions, we developed two simplified models that can approximate the phase-diagrams. The first simplified model gives us two simple scaling relationships which are in excellent agreement with the numerical simulations in the regime where both the silicon and the constraint are flowing plastically. The second simplified model which captures the dynamics of the problem and further reproduces the qualitative trends of the phase-diagrams. Our main conclusion is that lower axial growth may be achieved with relatively thin constraints as long as their yield stresses are high. This is an important finding for design considerations because it can save on overall volume requirements for a possible full-scale battery pack using silicon anode particles. 

%
%

\section*{Acknowledgements}
J.C., S.J.C., and A.G.  acknowledge support from the EPSRC through Grant No. EP/I017070/1. C.P.P. acknowledges support from the EPSRC through Grant No. EP/I01702X/1. A.G. is a Wolfson/Royal Society Merit Award Holder and acknowledges support from a Reintegration Grant under EC Framework VII. J.C. thanks Prof. Allan Bower (Brown University), Dr Giovanna Bucci (MIT), and Prof. Jianmin Qu (Northwestern University) for clarifications in their papers.

\bibliographystyle{unsrt}
\bibliography{refs}

\end{document}